\documentclass[aps,prd,nofootinbib,showpacs,amssymb,twocolumn]{revtex4}
\usepackage{amsmath,amsfonts}
\usepackage{epsfig,subfigure}
\usepackage{ifpdf}
\usepackage{hyperref}
\usepackage{amsmath}
\usepackage{graphicx}
\usepackage{color}
\usepackage{natbib}
\usepackage{multirow}

\newcommand{\be}{\begin{equation}}
\newcommand{\ee}{\end{equation}}
\newcommand{\bea}{\begin{eqnarray}}
\newcommand{\eea}{\end{eqnarray}}

\newcommand\editremark[1]{ {\color{red} #1}}

\newcommand\hidetosubmit[1]{}
\renewcommand\hidetosubmit[1]{#1}
\newcommand\ForInternalReference[1]{}

\newcommand\unit[1]{\, {\rm #1}}

\newcommand\Y[1]{Y^{(#1)}}

\newcommand\qmstateproduct[2]{\left\langle#1|#2\right\rangle}

\begin{document}

\author{Hee-Suk Cho$^{1}$}
\email{chohs1439@pusan.ac.kr}

\author{Evan Ochsner$^{2}$}
\email{evano@gravity.phys.uwm.edu}

\author{Richard O'Shaughnessy$^{2}$}
\email{oshaughn@gravity.phys.uwm.edu}

\author{Chunglee Kim$^{3}$}

\author{Chang-Hwan Lee$^{1}$}

\affiliation{$^1$Department of Physics, Pusan National University, Busan 609-735, Korea}

\affiliation{$^2$Center for Gravitation and Cosmology, University of Wisconsin-Milwaukee, Milwaukee, WI 53211, USA}

\affiliation{$^3$Department of Physics, West Virginia University, PO Box 6315, Morgantown, WV 26505, USA}

\title{Gravitational waves from BH-NS binaries: Effective Fisher matrices and parameter estimation using higher harmonics}
\date{\today}

\begin{abstract}
Inspiralling  black hole-neutron star (BH-NS) binaries emit a complicated gravitational wave signature, produced by multiple
harmonics sourced by their strong local gravitational field and  further modulated by the orbital plane's precession.
Some features of this complex signal are easily accessible to ground-based interferometers (e.g., the rate of change of
frequency); others less so (e.g., the polarization content); and others unavailable (e.g., features of the signal out of band).
For this reason, an ambiguity function (a diagnostic of dissimilarity) between two such signals  varies on many parameter scales and ranges.  
In this paper, we present a method for computing an approximate, \emph{effective} Fisher matrix
from variations in the ambiguity function on physically
pertinent scales which depend on the relevant signal to noise ratio.  
As a concrete example, we explore how higher harmonics improve parameter measurement accuracy.  
As previous studies suggest, for our fiducial BH-NS binaries and for plausible signal amplitudes, we see that higher
harmonics at best marginally improve our ability to measure parameters.  
For non-precessing binaries, these Fisher matrices \emph{separate} into intrinsic (mass,
spin) and extrinsic (geometrical) parameters;  higher harmonics principally improve our knowledge about the line of
sight.
For the precessing binaries, the extra information provided by higher harmonics is  distributed across several
parameters.  
  We provide concrete estimates for measurement accuracy,  using coordinates adapted to the precession cone
in the detector's sensitive band.        
\end{abstract} 


\maketitle

\ForInternalReference{
\editremark{Reminders}

--- TAKE OUT ALL DETAILS OF SMALLEST SCALE: Too many tables!  Too much to check against subtle numerical effects! ONLY
KEEP FOR LEADING-ORDER, non-spinning  for now  (possibly return to it  later)

 - global fit check: need rms errors for global 

- just for fun: 3d contour plot of the aligned-spin error ellipsoid, with and without higher harmonics

--- do we want to go beyond a single-point fit, because the ellipsoid is so long?  Have we walked down the ellipsoid to
insure the functional form is the same along it?  [basically, the eigenvalue associated with $\chi$ corresponds to a
  $P=0.99$ contour that extends over all $\chi$.]

--- very tricky, because the contouor is so extended in one direction: show raw data samples.  

--- NEED MORE POINTS IN $\chi$

- recheck covariances in quadratic limit...looks weird.

}

\section{introduction}
Ground based gravitational wave detector networks (notably LIGO \cite{Abb04} and Virgo
\cite{virgo})  are analyzing results of design-sensitivity searches
for the signals expected from the inspiral and merger of double compact binaries.
\cite{Abb08,Abb09}.
For the lowest-mass compact binaries $M=m_1+m_2\le 16 M_\odot$, the response of the detector to a binary merger  with arbitrary masses,
spins, and even eccentricity is well understood, particularly given the detectors' limited and low sensitive frequency band
\cite{Buo03,Buo04,Dam04,Pan04,Kon05,Buo05,Kon06,Tes07,Han08,Hin08,Aru09,Buo09}.  
Though this complicated signal  encodes all information about the binary's spacetime \cite{Rya95}, the amount of
accessible information depends on signal strength (or signal-to-noise ratio) \cite{Jar12}.  Strong signals permit high-precision
tests of general relativity; fainter signals allow high-precision constraints on some binary parameters;
 while very faint, short signals may only constrain the binary's mass.
Qualitatively speaking, we can distinguish two configurations if they are separated by contours $1-P\gtrsim 1/\rho^2$,
where $P$ (defined in Sec.~\ref{sec:Distinguish}) is the (normalized) ambiguity function or ``overlap'' and $\rho$ is the signal-to-noise ratio (SNR).  

As higher-order corrections and new physics are added to our models for gravitational-wave signals, the functional 
dependence on various parameters (such as masses, spins and orientation angles) in the model grows in complexity.
On scales $\simeq 1/\rho^2$, for astrophysically plausible $\rho$, the ambiguity function is generally smooth. However, 
it may have more complicated fine-scale structure which may not be detectable for expected signal strengths.
The Fisher matrix approach to estimating parameter errors is based on differentiating a waveform with respect to its 
parameters. These derivatives are defined in an infinitesimal patch of parameter space, and are thus measuring 
fine-scale structure, which could potentially be misleading about larger-scale, observable trends.
This point is illustrated by Fig.~\ref{fig2}, where we fit a quadratic through the ambiguity 
function. 
The ambiguity function changes shape, so fits to small ($P > 0.999$) and large ($P > 0.99$) regions of parameter space
would give rather different estimates of posterior widths and thus parameter accuracy. 
Similarly, if the standard Fisher matrix were calculated via finite difference, 
step sizes on these scales would give different results, with the \emph{smaller} step size giving a 
misleading estimate of parameter accuracy for a signal of expected strength.

%
%

%
In this paper we propose a simple effective procedure to identify relevant scales and parameter correlations, construct
suitable ``effective Fisher matrices'', and estimate ambiguity functions at low but nontrivial signal to noise ratio.
To demonstrate this technique, we examine the signal from selected black hole-neutron star (BH-NS) binaries, described
in Section \ref{sec:Sim}.  In this paper we use all available knowledge about the (post Newtonian) waveform, adopting a complete model for the adiabatic
quasicircular inspiral of precessing BH-NS binaries.   In particular, we employ all available
harmonics and amplitude corrections, introducing small but non-negligible changes to the ambiguity function.  
In Section \ref{sec:Distinguish} we introduce our unconventional effective approach to the local ambiguity
function.    
Using those tools, in Section \ref{sec:Results} we construct and approximate the overlap for signals similar
to each reference binary.   Motivated by parameter estimation, we provide explicit expressions for the Fisher matrix,
correlations, and marginalized uncertainties for each configuration.   
For our fiducial configurations, higher harmonics principally allow us to improve our knowledge of the
binary \emph{orientation}, providing fairly little additional information about intrinsic parameters for the amplitude scales of
immediate astrophysical interest.    

Our results are complicated by coordinate-dependent effects, notably  extreme sensitivity to the reference frequency at
which parameters are specified.  We show the choice of reference frequency can reduce (or introduce) fine-scale
structure into the ambiguity function, similar to the effect of higher harmonics.  
Our effective approach can partially compensate for ill-chosen coordinates, such as the coalescence phase
or initial spins.   To reduce but not completely eliminate these systematic effects,   we express our results using
parameters specified near the peak sensitivity of the detector (here, 100 {\rm Hz}).

\subsection{Context and prior work}

Several studies of gravitational wave detection from merging binaries have employed amplitude-corrected waveforms and
higher harmonics.    
Investigations of space-based interferometers, such as the Laser Interferometer Space Antenna (LISA),  
have historically used complete signal models, accounting for
both spin and precession \cite{Por08}.  As higher harmonics have a small effect, however, most previous studies of ground-based
interferometers have omitted them, emphasizing spin.  When included, higher harmonics were explored alone for
non-precessing signals.
   Higher harmonics can allow detection of signals otherwise inaccessible
due to the detector's limited bandwidth  \cite{Aru07,Bro07}.  The relative amplitudes of higher harmonics
can probe astrophysical mechanisms for generating non-circularity
\cite{Key10}.
Finally, higher harmonics (and precession) are well-known to break degeneracies and improve sky localization, particularly
for LISA \cite{Lan06,Kle09,Lan11}.  
%

Several authors have explored the local ambiguity function ``beyond the Fisher matrix'', including higher-order
correlation functions \cite{Val08,Vit10} and projection effects due to the local shape of the signal manifold \cite{Val11}.
These methods still use explicit derivatives of the ambiguity function (via explicit derivatives of the signal) to
construct their series approximations.

The Fisher matrix is often nearly or exactly singular, making inversion numerically challenging.
Several authors have pointed out that a singular value implies an unconstrained parameter, limited only by the prior;
see, e.g., Vallisneri \cite{Val08}.
In many cases, including those singular values addressed in the text, the singular value corresponds to a \emph{bounded}
parameter (e.g., an angle).  The singular value simply indicates that parameter cannot be measured. In the phenomenological limit described in this paper,   precisely zero eigenvalues never occur, unless a parameter is constrained by
symmetry. 

Our goal in this work is to understand the typical shape of the posterior $p(\lambda|n,\lambda_0)$ for $n$ a noise realization and $\lambda,\lambda_0$ coordinates in the signal space.   Using one notion of ``typical'' would produce an \emph{average} posterior over all noise realizations.   Such an average posterior, however, could be slightly wider than the posterior from any given noise realization.  Instead, in this work we attempt to characterize the typical shape of any \emph{one} noise realization.  To do so, in effect we ``transport'' each posterior so their peak likelihoods lie at the same point in parameter space.  In practice, our procedure amounts to ignoring noise-realization-dependent changes to the posterior.

\section{Simulations}
\label{sec:Sim}
\subsection{Amplitude-corrected precessing waveform}

In this paper we construct the post Newtonian (pN) gravitational wave signal from a BH-NS binary using the
\texttt{lalsimulation SpinTaylorT4} code \cite{lal}, which is an implementation based on the waveforms
described in \cite{Buo03,Buo04}.   This time-domain code solves for the orbital dynamics of
an adiabatic, quasicircular inspiralling binary by using the so-called TaylorT4 method 
(see \cite{Buo09} for an explanation of this and similar methods) of evolving the 
orbital phase and frequency supplemented with precession equations to 
track the motion of the spins and orbital plane \cite{Kid95}. 
The orbital phase and frequency evolution includes non-spinning corrections to 3.5pN order and 
spin corrections to 2pN order. The precession equations are given to 2pN order.
This binary evolution is terminated prior to merger, either when it reaches the ``minimum energy circular orbit'', 
or when the orbital frequency ceases to increase monotonically.

At each time, the values of the gravitational wave polarizations measured by a distant observer
can be constructed from the orbital phase, orbital frequency and the orientations of the spins 
and orbital plane. We can construct either the commonly used ``restricted'' (i.e. leading-order)
polarizations which contain only the dominant second harmonic of the orbital phase, 
or we can construct amplitude-corrected polarizations which contain terms that oscillate at other 
harmonics of the orbital phase (and also higher-order corrections to the second harmonic). 
Expressions for the polarizations valid for quasi-circular, precessing binaries 
are currently known to 1.5pN order  \cite{Aru09,Kid95,Wil96}.    Throughout this work, when we 
refer to amplitude-corrected waveforms, we mean that we use the 1.5pN accurate polarizations.

\subsection{Simulation coordinates}

\begin{figure}[t]
\includegraphics[width=8cm,height=6cm]{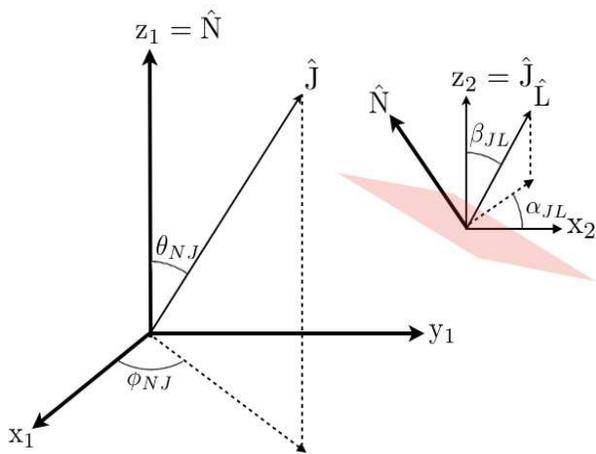}
\caption{{\bf Coordinate system for the precessing binary}. The left coordinate corresponds to the conventional GW radiation frame. $\theta_{NJ}$ ($\phi_{NJ}$) is a polar (azimuthal) angle of the total angular momentum ($J$) with respect to the radiation vector ($N$). In the left coordinate $\beta_{JL}$ ($\alpha_{JL}$) is a polar (azimuthal) angle of the orbital angular momentum ($L$) with respect to the total angular momentum ($J$). In the right coordinate, $N$, $J$, and ${\rm x_2}$ are coplanar and the shaded region indicates the orbital plane.\label{fig1}} 
\end{figure}

For the precessing binaries, LIGO-scale studies have been complicated by poor choice of coordinates, associated with the start of the waveform. The waveform generation code of the standard LIGO algorithm defines all the geometrical parameter values at the initial frequency (40 Hz for the initial LIGO and 10 Hz for the advanced LIGO), and evolve the binary system to get the full waveforms. Specifically, the orbit is described at some point, including the spin, orbital angular momentum vector, and the orbital phase. 
By contrast, the detector is more sensitive to higher frequencies. Allowing for the decreasing signal strength with frequency \cite{Bro12}, the detector is most sensitive to the instantaneous binary configuration at 100 Hz (initial LIGO) and 40 Hz (advanced LIGO, e.g., see Fig.~2 in \cite{Bro12}).  
Motivated by this fact, we choose the reference frequency ($f_{\rm ref}$), at which the instantaneous orientations of the spins and orbital plane are defined, to be 100 Hz. 

One significant effect introduced by setting the reference frequency at 100 Hz is related to the orbital phase. The ambiguity function is dependent on how we choose the reference frequency. In appendix B, we describe an example and illustrate the significant effects in detail; see also Fig.~\ref{fig2}. 

Furthermore, following Brown et al.\cite{Bro12}, we define the geometrical parameters to be angles between the radiation
vector $\hat{N}$, total angular momentum axis $\hat{J}$ and orbital axis $\hat{L}$ as in Fig.~\ref{fig1}.   
For comparison, other conventions specify some parameters using the line of sight, a vector $-\hat{N}$ pointing from our
detector to the binary.  

For non-spinning or aligned-spin binaries, the total angular momentum is parallel to the orbital angular momentum and the orbital axis is fixed.  In effect, the conventional radiation frame is equivalent to the geometrical frame.

\subsection{Fiducial simulations and local coordinates}
In the case of a non-spinning binary, the binary is specified by 9 parameters. In this work, we choose masses ($m_1,
m_2$) as intrinsic parameters, the polar (inclination) and azimuthal (polarization) angles ($\iota, \psi$) of the orbital axis with respect to the radiation vector and an orbital phase
$\phi$ as extrinsic parameters.   Because we  maximize the  ambiguity function over the polarization, we need not take
this parameter into explicit account henceforth.     Remaining parameters are, the distance to the detector, sky position (two angles), and the
coalescence time. The fiducial values of parameters are summarized in Table~\ref{tab1}. Mass components ($M=m_1+m_2$)
can be expressed by the symmetric mass ratio $\eta=m_1 m_2 / M^2$ and chirp mass $M_{\rm c}=M \eta^{3/5}$, we adopt these
parameters in this work. 

If the NS spin is assumed to be 0, the aligned-spin binary is specified by 10 parameters. 9 parameters are the same as the non-spinning case and the additional intrinsic parameter is dimensionless BH spin parameter $\chi$. The fiducial values are also summarized in Table~\ref{tab1}.  

\begin{table}[!]
\begin{tabular}{c|ccccc|cc }
parameter&$m_1$&$m_2$& $\iota$& $\phi$&$\chi$& $M_{\rm c}$  & $\eta$ \\
\hline
non-spinning&10&1.4&$\pi/4$&0.0 &0.0& 2.994 & 0.1077\\
aligned-spin&10&1.4&$\pi/4$&0.0 &1.0& 2.994 & 0.1077\\

\end{tabular}
\caption{{\bf Fiducial  parameters for the non-spinning and aligned-spin binaries.} We adopt the chirp mass $M_{\rm c}$ and symmetric mass ratio $\eta$ instead of individual masses. 
The orbital phase is defined at 100 Hz.\label{tab1}}
\end{table}

The waveform of the precessing binary can be defined by 12 parameters if the NS spin is assumed to be 0. In this work,
we consider $\eta$, $M_{\rm c}$, BH spin $\chi$, and the opening angle $\beta_{JL}$ of the precessing cone as intrinsic
parameters, $\alpha_{JL}$, $\theta_{NJ}$, $\phi_{NJ}$, and the orbital phase $\phi$ as extrinsic parameters.   Because
we 
maximize over the polarization angle $\psi$, the parameter $\phi_{NJ}$ is eliminated from further consideration.   Remaining parameters are the distance, sky position (two angles), the coalescence time. 
Throughout this paper the units are solar masses  (for $M_{\rm c}$); radians (for angles); or
the natural dimensionless units  (for $\eta,\chi$).

Motivated by \cite{Bro12}, we adopt a challenging reference configuration, where the polarization 
along the line of sight oscillates between circularly polarized ($L$ along $N$) and linearly 
polarized ($L$ perpendicular to $N$).   Furthermore, to explore the extent to which higher-order 
harmonics allow measurement of parameters that only weakly impact the signal, we consider two 
possible sets of initial conditions for $L$ along its precession cone. The fiducial values of the 
parameters are summarized in Table~\ref{tab2}. For case1, the orbital axis is perpendicular to the 
radiation vector at 100 Hz, for case2 it is parallel to the radiation vector at 100 Hz. All the 
parameter values are the same between both cases except for $\alpha_{JL}$.

\begin{table}[!]
\begin{tabular}{c|ccccccc|c }
parameter&$m_1$&$m_2$& $\chi$ & $\beta_{JL}$ &  $\theta_{NJ}$ & $\alpha_{JL}$ & $\phi$& configuration\\
\hline
case1&10&1.4&1.0&$\pi/4$&$\pi/4$&0.0&0.0&$N \perp L$\\
case2&10&1.4&1.0&$\pi/4$&$\pi/4$&$\pi$&0.0&$N \parallel L$
\end{tabular}
\caption{{\bf Fiducial  parameters for the precessing binary}: We adopt the chirp mass $M_{\rm c}$ and symmetric mass ratio $\eta$ instead of individual masses. 
All the extrinsic parameters are defined at 100 Hz. 
For the case1, $L$ is perpendicular to $N$,  and along for the case2 at 100 Hz.\label{tab2}}
\end{table}

\subsection{Fiducial network}
We assume two identical interferometers placed perpendicular to the incident signal, which is the optimal sky position
of the source. We also assume the two interferometers are oriented by $\pi/4$ related to one another, giving comparable
sensitivity to both polarizations. For the incident waveforms, we assume a zero noise limit to understand how similar the signals are. While not realistic, they avoid introducing complexity of the signal due to the source sky position.

\section{distinguishing simulations}
\label{sec:Distinguish}
\subsection{Ambiguity function}
In this work, we reorganize the two projections of the strain tensor $h_+ = e_+^{ab}h_{ab}/2$ and
$h_\times=h_{ab}e_{\times,ab}/2$ into a complex function:
\be
h(t) \equiv h_+(t) + i h_\times(t).
\ee
We coherently compare a fiducial signal $h_0(t,\lambda_0)$, where $\lambda_0$ indicates a fiducial source parameter set, to a nearby signal $h(t,\lambda)$, with parameters $\lambda$, by a complex overlap \cite{Osh12}
\be
\label{eq.complexoverlap}
\langle h_0 | h \rangle \equiv 2  \int^{\infty}_{-\infty}  \frac{df}{S_n(f)}[\tilde{h}_0(f)\tilde{h}(f)^*], 
\ee
where $\tilde{h}(f)$ is the Fourier transform of $h(t)$ and $S_n(f)$ is a detector strain noise power spectrum. For simplicity, we adopt a semianalytic initial LIGO 
sensitivity \cite{Dam01,Aji09}. 
As pointed out by \cite{Osh12}, this complex-valued expression characterizes the ability of a network to distinguish signals.
The real part of the complex overlap corresponds to  a linear sum of the conventional real overlaps of the two gravitational wave
polarizations:
\begin{eqnarray}
\label{eq.realoverlap}
\text{Re}\langle h_0 | h \rangle  = (h_{0,+}|h_+) + (h_{0,\times}|h_{\times}),
\end{eqnarray}
where $(h_0|h)$ indicates the conventional overlap of two real functions defined by
\be \label{eq.conventionaloverlap}
( h_0 | h) \equiv 4  \int^{\infty}_0  \frac{df}{S_n(f)}{\rm Re}[\tilde{h_0}(f)\tilde{h}(f)^*] .
\ee
In appendix A, we summarize the differences between the real and complex overlaps. 

We note that a change of the polarization angle, $\psi$, simple causes a rotation of the argument 
of the complex wave strain function, $h(\psi) = e^{-2i\psi}h(\psi=0)$. Thus it is trivial to find the 
value of $\psi$ which makes the complex overlap purely real, so that
\bea {\rm Im}\langle h_0|h' \rangle&=&0,\\ \nonumber
         {\rm Re }\langle h_0|h' \rangle&=&(h_{0,+}|h'_+)+(h_{0,{\times}}|h'_{\times}),
\eea         
and the value of the complex overlap maximized over polarization angle $\psi$ is simply
\be\label{eq.max}
{\rm max}_{\psi}\ {\rm Re}\langle h_0|h \rangle=|\langle h_0|h \rangle|.
\ee
The complex overlap (like the real-valued overlap) can also be  maximized over the coalescence 
time $t_{\rm c}$ via an inverse Fourier transform as described in \cite{All05}. In particular, one 
uses the fact that 
\be 
\tilde{h}(t_{\rm c}=t) = \tilde{h}(t_{\rm c}=0) e^{-2 \pi i f t}
\ee
and notes that the inverse Fourier transform of the complex overlap integrand in 
Eq.~(\ref{eq.complexoverlap}) will compute the complex overlap for all possible coalescence times
of $h$ at once
\be
\langle h_0 | h(t_{\rm c}=t) \rangle \equiv 2  \int^{\infty}_{-\infty}  \frac{df}{S_n(f)}[\tilde{h_0}(f)\tilde{h}(f)^*] e^{2 \pi i f t}.
\ee

The (normalized) ambiguity function $P(\lambda_0,\lambda)$ between two waveforms $h_0(t,\lambda_0)$ and $h(t,\lambda)$ is then 
defined as the complex overlap maximized over
polarization angle and coalescence time,  
\be\label{eq.ambiguity}
P(\lambda_0,\lambda)\equiv {\rm max}_{t_{\rm c},\psi}{|\langle h_0 | h \rangle |  \over  \sqrt{\langle h_0 | h_0 \rangle \langle h | h \rangle }} .
\ee
Unless otherwise noted,  all overlaps are maximized in time and polarization. This is different from maximizing over orbital phase $\phi$; see appendix A and B.

\subsection{Likelihood}
The detector noise $N(t)$ is assumed to be a stationary and Gaussian process. Given the detector output $S(t)=H(t,\lambda_0)+N(t)$ representing a real-valued signal in real-valued noise,  the probability for the noise to have some realization $N_0$ is  \cite{Fin92}
\be
p(N=N_0) \propto e^{-(N_0 | N_0 )/2}.
\ee
The posterior probability that the gravitational wave signal is characterized by the parameters $\lambda$, can be expressed by 
$p(\lambda|S) \propto p(\lambda)L(S|\lambda)$, where $p(\lambda)$ is the prior probability that the signal is characterized by $\lambda$, $L(S|\lambda)$ is the likelihood, which can be written by  \cite{Fin92}
\be\label{eq.preal}
L(S|\lambda) =C \times {\rm exp}\bigg[-\frac{1}2( S-H(\lambda)|S-H(\lambda))\bigg],
\ee
where $C$ is a proportional factor which, for simplicity, we assume to be 1 in this work.

Since we consider the complex strain, by choosing the appropriate polarization angle we shall write the detector output for the detector 1 and 2. 
\be\label{eq.sreal}
S_1=H_++N_1, \ \ \ S_2=H_\times+N_2,
\ee
also
\bea\label{eq.scomplex}
s=S_1+iS_2, \ \ \ h_0=H_++iH_\times, \ \ \ n_0=N_1+iN_2.
\eea
The probability for the noise to have both realizations $N_1$ and $N_2$ is
\bea\label{eq.pn}
p(N_1,N_2)&\propto& e^{-\langle N_1 | N_1 \rangle/2} e^{-\langle N_2 | N_2 \rangle/2}\\  \nonumber
                &=&e^{-{\rm Re}\langle N_1+iN_2 | N_1+iN_2 \rangle/2}.
\eea
Finally, using Eqs.~(\ref{eq.sreal} - \ref{eq.pn}), Eq.~(\ref{eq.preal}) can be 
expressed by the \emph{complex} signals:
\be
\label{eq.pcomplex}
L(s|\lambda) = {\rm exp}\bigg[-\frac{1}2 {\rm Re} \langle s-h(\lambda)|s-h(\lambda) \rangle \bigg].
\ee
Substituting $s=h_0+n_0$ into this equation \cite{Cut07}, the likelihood is
\bea
L&=&{\rm exp} \bigg[ -\frac{1}2 {\rm Re}\langle n_0+h_0-h|n_0+h_0-h \rangle \bigg]  \\ \nonumber
       &=&{\rm exp} \bigg[ -\frac{1}2 {\rm Re}\{\langle h_0-h|h_0-h\rangle +2 \langle n_0|h_0-h \rangle \\ \nonumber
       && \  \ \ \ \ \ \  + \langle n_0|n_0 \rangle\} \bigg],
\eea
where the second and third terms in the a square bracket depend on the noise realization. They shift the position of the maximum likelihood but only weakly change the shape of the likelihood curve.  In the limit of high SNR, these noise-dependent terms can be neglected, so,
\bea \label{eq.p1}
L&=&{\rm exp} \bigg[ -\frac{1}2 {\rm Re}\langle h_0-h|h_0-h\rangle \bigg]\\   \nonumber
&=&{\rm exp} \bigg[ -\frac{1}2\{\langle h_0|h_0 \rangle+\langle h|h\rangle-2{\rm Re}\langle h_0|h\rangle\} \bigg].
\eea
This equation corresponds to the case where two detectors are placed to have the maximum response to the incident two polarizations [for the detector placement, see Section \ref{sec:Sim} D]. While, Eq.~(\ref{eq.preal}) corresponds to one detector response to one polarization; see appendix A.

Using Eqs.~(\ref{eq.max}) and (\ref{eq.ambiguity}), and the SNR defined by $\rho^2 =\langle h_0|h_0 \rangle=\langle h|h \rangle$, 
the log likelihood can be expressed by our complex overlap convention:
\be \label{eq.lnp}
\ln L= -\rho^2(1-P)
\ee
where we assume the same strength for both signals.

For a given log likelihood, the scale of interest of the ambiguity function depends on the signal strength $\rho^2$:
\be\label{eq.rho}
1-P \le \frac{1}{\rho^2}.
\ee
By approximately identifying the $L\simeq 1/e$ surface of the likelihood, this condition allows us to estimate the set of
parameters $\lambda$ which cannot be distinguished from $\lambda_0$ with a signal amplitude of $\rho$ using a signal
model and noise curve that produces an overlap $P(\lambda_0,\lambda)$.\footnote{More properly, the probability 
  $F(L_0)=\int_{L>L_0}  p(x) d\lambda$ of having likelihood $>L_o$ lets us create a confidence volume for any target
  confidence level.   Because the probability $p\equiv F(L_o)$  depends sensitively on $L$, we anticipate the edge of this confidence
  interval will depend weakly (e.g., as  $F^{-1}[p]\propto \sqrt{\ln p}$) on the precise probability used to define the threshold.}
\subsection{Fisher matrix}

If $\lambda$ is close to $\lambda_0$, we can write $h_0-h$ to the first order in the error $\Delta \lambda_i \equiv \lambda_0 - \lambda$ 
\be
h_0-h \sim {\partial h \over \partial \lambda_i} \Delta \lambda_i.
\ee
So, in the limit of high SNR, the likelihood [Eq.~(\ref{eq.p1})] is given as $L = {\rm exp}(-\Gamma_{ij} \Delta \lambda_i \Delta \lambda_j/2) $, where $\Gamma_{ij}$ is 
\be \label{eq.analyticfisher}
\Gamma_{ij}={\rm Re}\bigg\langle {\partial h \over \partial \lambda_i} \bigg | {\partial h \over \partial \lambda_j} \bigg \rangle.
\ee
This definition is analogous to the standard Fisher matrix, except that it is derived from the complex 
overlap and therefore contains information about both polarizations.
If we assume that the prior $p(\lambda)$ is uniform, the parameter estimation errors $\Delta \lambda_i$ (i. e., the posterior probability density function) can be expressed by the Gaussian distribution
\be  \label{eq.posterior}
p(\Delta \lambda_i)=N e^{-\Gamma_{ij} \Delta \lambda_i \Delta \lambda_j/2}
\ee
where $N=\sqrt{{\rm det}(\Gamma /2\pi)}$ is the corresponding normalization factor.

Using another expression relating the Fisher matrix to the log likelihood \cite{Jar94,Val08} and Eq.~(\ref{eq.lnp}), we can write
\bea
\Gamma_{ij}&=&-{\partial^2 \ln L(\lambda) \over \partial \lambda_{i} \partial \lambda_{j}}\bigg|_{\lambda_i=\lambda_j=\lambda_0}\\ \nonumber
                          &=&\rho^2  {\partial^2 (1-P) \over \partial \lambda_{i} \partial \lambda_{j}}\bigg|_{\lambda_i=\lambda_j=\lambda_0}=\rho^2 \hat{\Gamma}_{ij},
\eea
where $\lambda_0$ is the fiducial value of source parameter. Here, we define the normalized Fisher matrix $\hat{\Gamma}_{ij}$.

For Gaussian noise and high SNR, the inverse of the Fisher matrix is the covariance matrix ($\Sigma_{ij}$) of parameter
errors. The measurement error ($\sigma_i$) of each parameter and correlation coefficient ($c_{ij}$) between two
parameters are defined as
\be
\sigma_i=\sqrt{\Sigma_{ii}}, \ \ \ c_{ij}={\Sigma_{ij} \over \sqrt{\Sigma_{ii}\Sigma_{jj}}},
\ee
The correlation coefficients $c_{ij}$ are $\rho$-independent but often sensitive to small changes in $\hat{\Gamma}$.
Conversely, the  measurement error is inversely proportional to $\rho$.  For the purposes of illustration, we adopt
$\rho=10$ whenever we calculate $\sigma_i$.


\subsection{Relevant scales and effective approach}

The Fisher matrix formally involves derivatives, i.e., infinitesimal variations of a parameter $d \lambda$. In this work, we 
compute an effective Fisher matrix by considering finite variations $\delta \lambda$ on scales which give 
physically observable changes to the ambiguity function, $P$.

To understand the variability on multiple scales we plot a one-dimensional ambiguity function of $M_{\rm c}$  for the 
leading-order amplitude, non-spinning binary in Fig.~\ref{fig2}. In this figure, the ambiguity function $P$ is calculated via 
Eq.~(\ref{eq.ambiguity}), changing only $M_{\rm c}$ and fixing all other parameters to be the same for both signals.
For comparison, we plot quadratic fits\footnote{Since the posterior function is a normal distribution in the limit of high SNR (see Eq.~(\ref{eq.posterior})), a quadratic fitting function usually best fits the log likelihood function with a flat prior.} to the ambiguity curve at different scales of  $P>0.99$ and $P>0.999$. These are the scales of interest for signals with strength $\rho^2 \sim 10^2$ or $10^3$ (see, Eq.~(\ref{eq.rho})). 
[Note that Fig.~\ref{fig2} is computed with the reference frequency at 40 Hz. For our results, except for this figure,
  we choose the reference frequency at 100 Hz. The structure illustrated here is present but  less significant at 100 Hz; see appendix B.]

The shape of the ambiguity function has structure on multiple scales. 
The neighborhood of $P>0.999$ suggests a much sharper peak than what is seen at the $P>0.99$ scale.
A Fisher matrix computed from formal waveform parameter derivatives (defined in the limit $ d \lambda \rightarrow 0 $)
or finite difference such that $P(\lambda_0, \lambda_0 + \delta \lambda) \gtrsim 0.999$ can be overly optimistic about how
well $\lambda$ can be measured for a signal with $\rho^2 \sim 10^2$.

Therefore, we wish to define an effective Fisher matrix from the curvature of the ambiguity function 
on the scales of interest. For example, in the case of two parameters, the fitting function $P^*$ is
\be
\label{eq.2Dfit}
P^*=P_{\rm max} + p_1 \delta \lambda^2_1+ p_2 \delta  \lambda^2_2 + p_{12}  \delta \lambda_1 \delta \lambda_2\ ,
\ee
where $p_i$ and $p_{ij}$ are  fitting coefficients and $P_{\rm max}=1$.
We calculate the effective Fisher matrix as
\be
(\hat{\Gamma}_{ij})_{\rm eff}=-{\partial^2 P^* \over \partial \lambda_i \partial \lambda_j}\ .
\ee

In some cases, especially when a parameter is poorly determined, the variation of the ambiguity function with a 
parameter may not be well-described by a quadratic. See, for example, the top panel of Fig.~\ref{fig5}. 
Therefore, we find it useful to employ an ``iterative'' procedure to find the parameters that are amenable to a quadratic fit.
For each parameter $x$, we compute the one-dimensional curve $P(x, x + \delta x)$ and fit it against 
$1-(\hat{\Gamma}_{xx})_{\rm eff}\delta x^2/2$. This determines the diagonal elements of the effective Fisher matrix, 
$(\hat{\Gamma}_{xx})_{\rm eff}$. We discard any parameters that are poorly fit by this quadratic 
(checked either ``by eye'' or a with quantitative threshold on the goodness of fit).
For the well-fit parameters, we determine the off-diagonal elements of the Fisher matrix by computing the two-dimensional 
surface $P(x,y;x+\delta x,y+\delta y)$ and fitting it to a function of the form like Eq.~(\ref{eq.2Dfit}) while using the values from the 
one-dimensional fits, i.e. $p_x = -(\hat{\Gamma}_{xx})_{\rm eff}/2$. We note this method is used primarily as a sanity check 
to identify any parameters which induce obviously non-quadratic variations in $P$.

Once we have identified the space of all reasonably quadratic parameters, the ambiguity function 
on that space can be approximated as
\be
P^*(\lambda_0, \lambda_0 + \delta \lambda) = 1 - (\hat{\Gamma}_{ij})_{\rm eff}\ \delta \lambda_i\ \delta \lambda_j / 2\ .
\ee
Rather than using the iterative procedure described above to find the elements of $ (\hat{\Gamma}_{ij})_{\rm eff}$ one at a time,
it is straightforward to use a standard least-squares fitting technique to simultaneously solve for all of the 
$(\hat{\Gamma}_{ij})_{\rm eff}$. This will also give a better global approximation to the ambiguity function 
than the iterative approach. As an example of the small yet noticeable differences between these two procedures, 
Table  \ref{tab:Fisher:AlignedSpin} compares the results for computing the effective Fisher matrix 
from both ``iterative'' and  ``simultaneous'' fits to the ambiguity function. Everywhere else in this work (Tables~\ref{tab3},
 \ref{tab4}, \ref{tab3:Alt}, \ref{tab:Fisher:AlignedSpin},  \ref{tab:Fisher:AlignedSpin:higher}, and \ref{tab:Fisher:Precessing})
 the effective Fisher matrix is computed by simultaneously fitting all coefficients.

In the cases where $P$ is not well-described by a quadratic (e.g., see Fig.~\ref{fig5}), we can adopt more complicated expressions to
characterize the functional dependence of $P$ when these parameters are varied, both in isolation and in correlation with
well-constrained variables.  
As a concrete example,  in the absence of higher harmonics the line of sight from the binary is both
\emph{weakly constrained} by observations and nearly \emph{separable} in $P(\lambda_0,\lambda)$ from other degrees of
freedom\footnote{Approximate separability of the line of sight from other degrees of freedom follows only in our well-chosen
  coordinates, where the binary configuration is specified at $100{\rm Hz}$. 
}.    Specifically, ignoring maximization in time and phase, the overlap
$P(\hat{N},\hat{N}')$ between any two lines of sight can be well-approximated by Eq.~(B1c) from \cite{Osh12}:
\begin{eqnarray}
\label{eq:AngularDependence:Leading}
P_{\rm angles} &\simeq & \frac{
  | e^{2 i \phi}Y_{2}(\theta)Y_2 (\theta')
 + 
  e^{-2 i  \phi}Y_{-2}(\theta)Y_{-2}(\theta')
   |
 }{
\sqrt{( |Y_{2}(\theta)|^2  + |Y_{-2}(\theta)|^2)(|Y_{2}(\theta')|^2+ |Y_{-2}(\theta')|^2)}
}  \nonumber\\
\end{eqnarray}
where we use the shorthand $Y_m\equiv \Y{-2}_{2m}(\hat{n})$ and similarly for $Y_m'$ to reduce superfluous subscripts
and where  we factor out the common $e^{im\phi}$ from $\Y{-2}_{lm}$.   This function has wide, nearly
flat extrema in $\hat{n}$ for each fixed $\hat{n}'$.   
%
On the other hand, in the absence of higher harmonics the line of sight has little impact on the waveform phase versus
time away from the orbital plane.   
We can therefore approximate the ambiguity function for $P>0.99$ in the top panel of Fig.~\ref{fig5} by
\begin{eqnarray}
\label{eq:OverlapIncludingNonquadraticAngles:Leading}
P \simeq P_{\rm angles}[1 - \frac{1}{2}(\hat{\Gamma}_{ab})_{\rm eff}  \delta \lambda_a \delta \lambda_b] - \Gamma_{aN} \delta
\lambda_a \delta \lambda_N\ ,
\end{eqnarray}
where the $N$ index varies over the line-of-sight parameters $(\theta,\phi)$ and the  $a,b$ 
indices vary over the other parameters and $\Gamma_{aN}\simeq 0$ for $a \notin N$. 
%
%
With higher harmonics, the functional form above [Eq.~(\ref{eq:OverlapIncludingNonquadraticAngles:Leading})]  is weakly
  perturbed by additional angular terms of the form
\begin{align}
P(\lambda_0,\lambda) &\simeq  P_{\rm angles} [1 - \frac{1}{2} (\hat{\Gamma}_{ab})_{\rm eff} \delta \lambda_a \delta \lambda_b] - \Gamma_{aN} \delta \lambda_a
\delta \lambda_N \nonumber \\
\label{eq:AngularDependence:Higher}
&- \frac{1}{2} (1-\cos \iota)G_{\phi\phi} (\phi-\phi_0)^2 
\\ & -\frac{1}{2}G_{cc}( \cos\iota -\cos\iota_0 ^2
\nonumber \\ & - G_{\phi c} (\phi-\phi_0)(\cos \iota - \cos \iota_0) \nonumber
\end{align}
where $G_{ab}$ is a matrix with $G_{\phi,c}\simeq 0$.     
This approximation both factors out the leading-order angular dependence and adds additional  angular terms with  parameter-dependent coefficients, designed to  correctly reproduce a $\phi$-independent
result when $\cos \iota \simeq 0$.  
Although these terms allow us to correctly reproduce the non-ellipsoidal contours seen in the bottom
panel of Fig.~\ref{fig5}, Tables~\ref{tab:Fisher:AlignedSpin} and \ref{tab:Fisher:AlignedSpin:higher} 
show that this complicated structure only marginally improves the overall fit compared to a purely
quadratic approximation $P\simeq 1-(\hat{\Gamma}_{ab})_{\rm eff} \delta \lambda_a \delta \lambda_b/2$.  
Fit parameters for this more complicated functional dependence are not presented here.  
%

We also considered a more general fit, treating $P_{\rm max}$ as a parameter.   While this parameterization has a 
significant aesthetic advantage -- its effective Fisher matrix is roughly scale-independent when $f_{\rm ref}\simeq
100 {\rm Hz}$ and agrees with analytic
calculations -- it systematically underestimates $P$ in the neighborhood of the maximum.  As both fits work well
globally,  we favor the simpler procedure and adopt $P_{\rm max}=1$ except for Table  \ref{tab3:Alt}.

\subsection{Comparing to standard Fisher matrix results}
Despite subtle differences associated with time domain versus frequency domain waveforms, the complex overlap, higher harmonics,  and the line of sight, our results
for the effective Fisher matrix are directly comparable to earlier results calculated with the stationary phase approximation  \cite{Poi95}. 
For example, for emission along the $\hat{z}$ axis, both the real and complex strain have the form\footnote{The two
  differ for $f<0$: the complex strain has $\tilde{h}(f<0,\hat{z})=0$, while the real strain has
  $\tilde{h}(-f)=\tilde{h}(f)^*$.}
$\tilde{h}=Ae^{-2i\Phi}$ for $f>0$.   As a result, the Fisher matrix  $\hat{\Gamma}_{ij}$  can be well-approximated by
an \emph{identical} average over frequency:
\begin{eqnarray}
\label{eq:AnalyticFisher:SPALeadingOrder}
\hat{\Gamma}_{ij} &=& 
\frac{\int_{-\infty}^{\infty} df |\tilde{h}|^2 (\partial_a \Psi)(\partial_b \Psi)/S_h}{\int_{-\infty}^{\infty} df |\tilde{h}|^2/S_h}
\end{eqnarray} 
where we neglect derivatives $\partial_a A$ as small compared to the leading-order phase dependence.  
Thus, \emph{each component} of our Fisher matrix must resemble previous results.
In fact, as the general definition [Eq.~(\ref{eq.analyticfisher})] suggests, each component of $\Gamma_{ab}$   (unlike
$\Sigma = \Gamma^{-1}$) depends
only on the local response to changing \emph{two} parameters $\lambda_a,\lambda_b$, no matter how many parameters exist
in the model.   Therefore, the Fisher matrix for an identical
model with more parameters will have, as a submatrix, the  Fisher matrix for the smaller model.  
By contrast, other methods for expressing uncertainty like the covariance $\Sigma$  depend simultaneously on \emph{all}
terms in $\Gamma$.   For low-mass binaries, the Fisher matrix $\Gamma$ is well-known to be poorly conditioned,
with eigenvalues spanning several orders of magnitude.  
We therefore  preferentially compare the \emph{component-by-component Fisher matrix}, rather than the covariances
$\Sigma$, when comparing results.
When presenting results, we provide several significant figures to insure all eigenvalues of $\Gamma$ remain positive-definite.

Our complex overlap maximizes over time and polarization.   The analytic Fisher matrix
calculated from the stationary phase approximation [Eq.~(\ref{eq:AnalyticFisher:SPALeadingOrder})] has time and phase as
parameters.   To account for maximizing over those parameters, we transform the full Fisher matrix $\Gamma$  to a
smaller-dimensional matrix which projects out those dimensions:
\begin{eqnarray}
\label{eq:review:Gaussian:DimensionalReduction}
(\hat{\Gamma}_{ab})_{\rm max}&=& \hat{\Gamma}_{ab} - \hat{\Gamma}_{aC} Q_{CD} \hat{\Gamma}_{Cb} \\
{} Q_{CD} &\equiv& [\hat{\Gamma}_{CD}] ^{-1}
\end{eqnarray}
where $C,D$ run over the $t,\phi$ variables and $a,b$ all other variables.   In these expressions, the matrix $Q$ is the inverse of the projection of $\hat{\Gamma}_{ab}$ into the $t,\phi$ subspace.

\subsection{Comparing to posteriors}
Standard parameter estimation techniques like Markov-Chain Monte Carlo produce samples of the full posterior probability distributions,
including postprocessed data products like one-dimensional standard deviations $\sigma_a$ and two-dimensional covariances
$c_{ij}$.  
For \emph{strong signals} with well-isolated probability distributions,  our one-dimensional standard deviations and
covariances are directly comparable, for identical binaries.  
For fainter signals with broad probability distributions, our results will describe part of the  posterior, in the
neighborhood of one extremum.

For brevity, we have explicitly eliminated two parameters -- event time and polarization -- and make no predictions
about any correlation including them.   We will revisit these parameters, along with asymmetric detector response, in a
subsequent publication.

\begin{figure}[!]
\centerline{
 \includegraphics[width=\columnwidth,height=6cm]{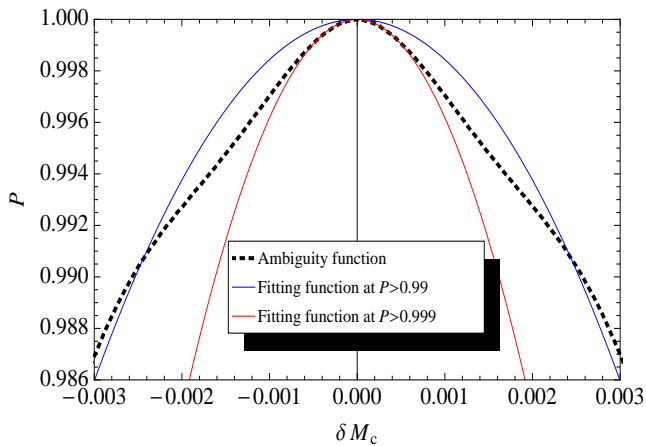} 
 }
\caption{\textbf{One dimensional ambiguity function for the non-spinning binary, showing variability on multiple
    scales}.  Dotted line shows $P$ calculated from Eq. \ref{eq.ambiguity} for the leading-order waveforms, where we
  compare our fiducial simulation with slightly offset analogs, changing just the chirp mass. For comparison, the two
  solid curves show quadratic fits to the ambiguity function, on scales of $P>0.99$ (blue) and $P<0.999$ (red). Note
  that this result is computed with the reference frequency at 40 Hz, a default choice that accentuates this scale
  dependence; see appendix B.
  In this paper we adopt a reference frequency at 100 Hz to mitigate the change in scale shown here; see Fig.~\ref{fig3} for the comparable result in that case.  \label{fig2}}    
\end{figure}

\begin{figure}
 \includegraphics[width=\columnwidth,height=6cm]{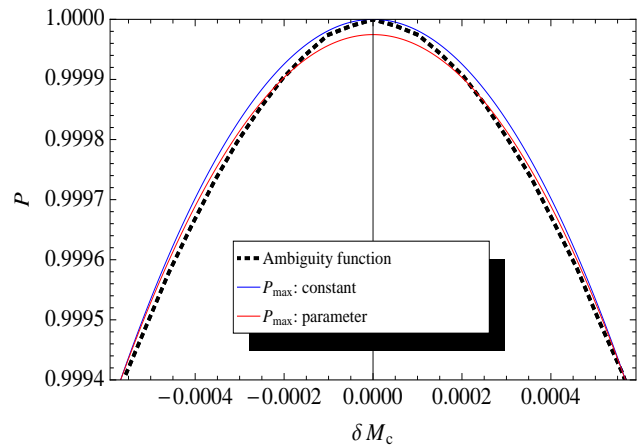} 
\caption{\textbf{Two methods for fitting the ambiguity function\label{fig3}}: Illustration
  of differences between our default
  fitting technique (blue curve), which  fixes $P_{\rm max}=1$, and an alternate fitting technique that lets $P_{\rm max}\ne
  1$, shown here for the $P>0.999$ scale (dotted curve).   These two methods produce comparable global fit \emph{functions}, but the fitting \emph{parameters} for $\Gamma$
  differ by tens of percent; see, e.g., Table~\ref{tab3:Alt}.     While numerical error can produce fluctuations in the overlap (e.g., due to
  insufficient sampling rate; we use $65536\unit{Hz}$), the change in shape shown here is resolved. 
}
\end{figure}



\begin{figure}[!]
\includegraphics[width=\columnwidth]{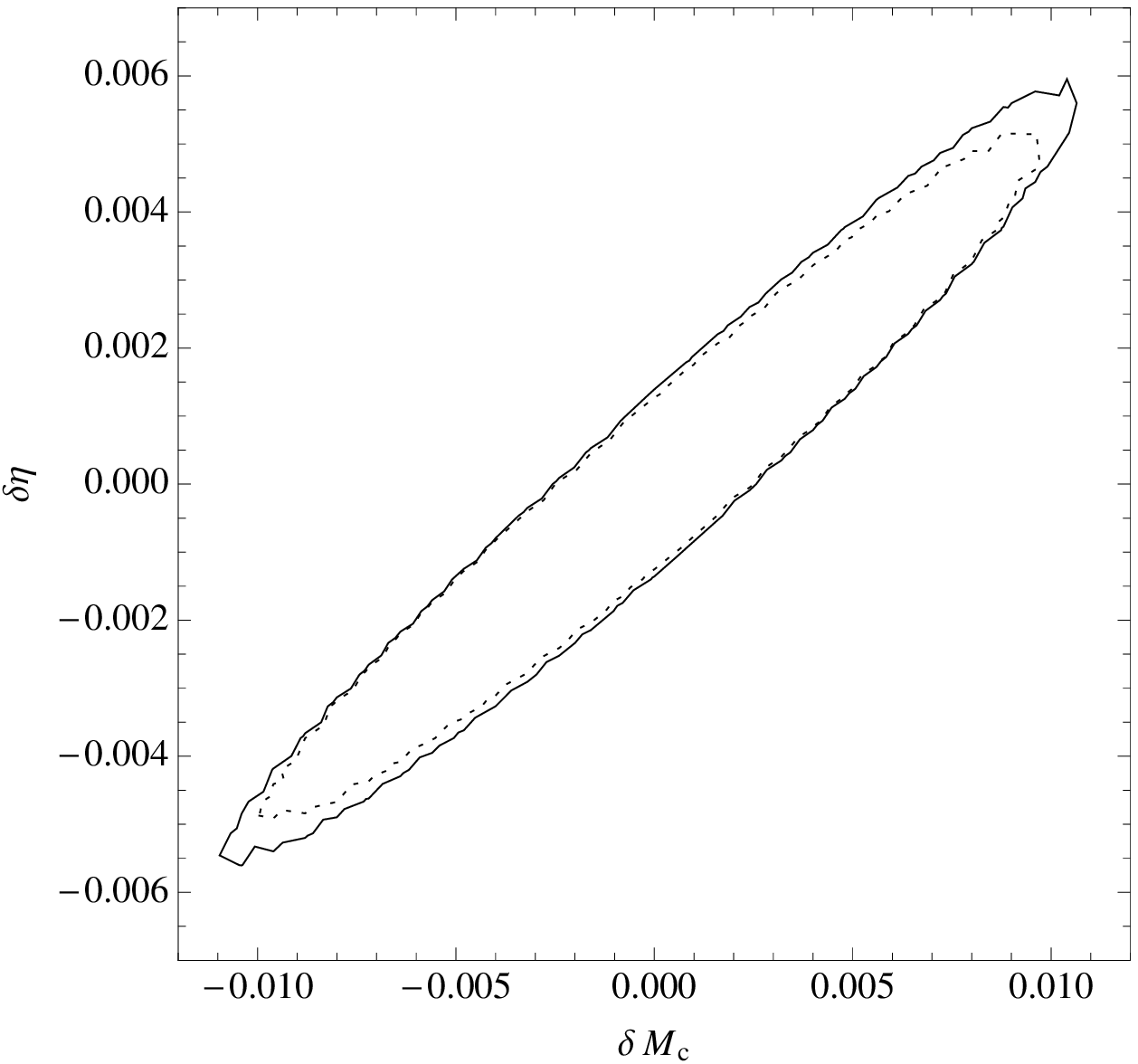}
\includegraphics[width=\columnwidth]{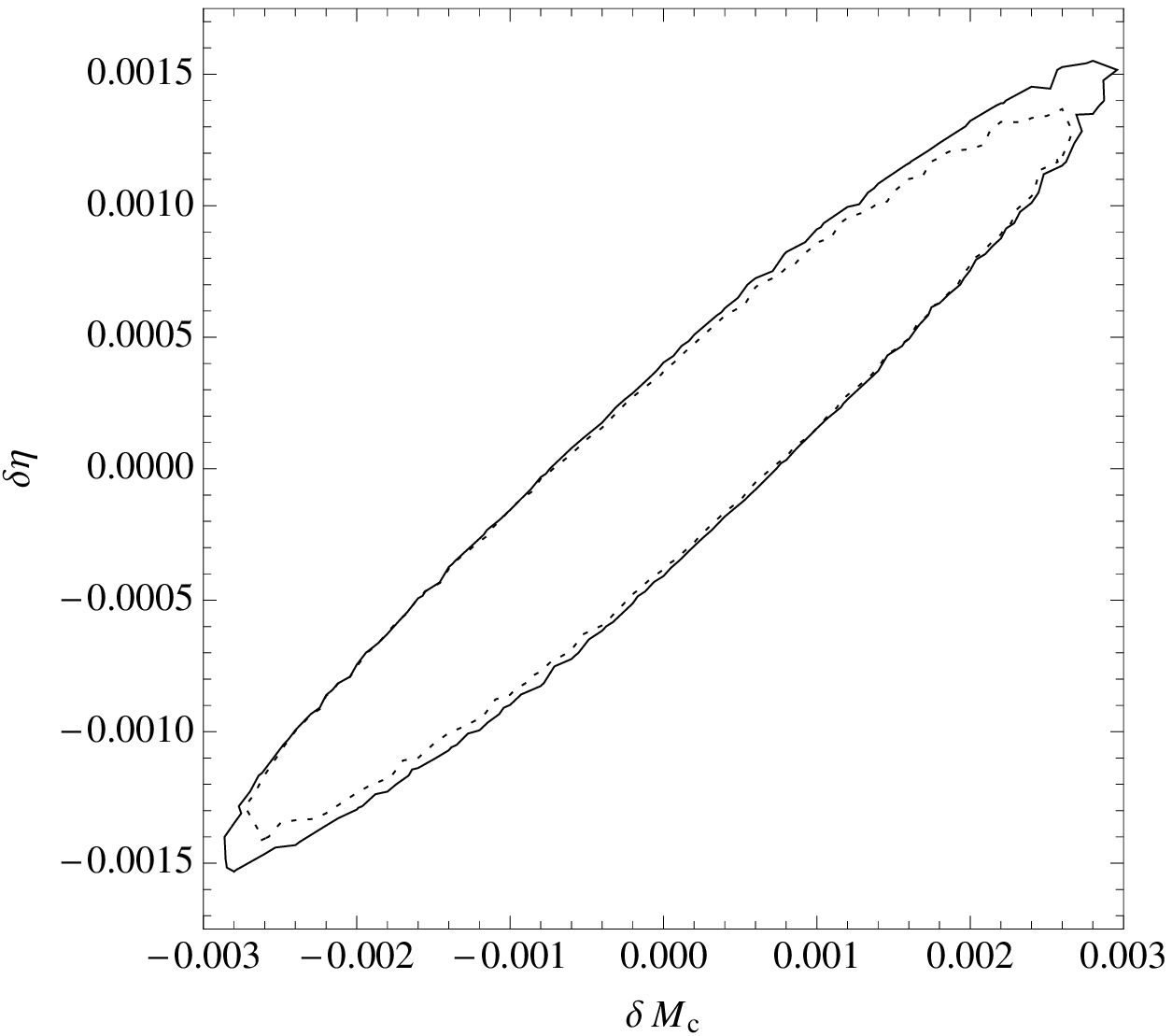}
\caption{{\bf Comparison of the ambiguity contours between the leading-order ({\bf solid line}) and higher-order ({\bf dotted line}) waveforms for the non-spinning binaries.}  ${\bf \delta \lambda}$ is defined by a difference from the fiducial value in the Table~\ref{tab2}. The lines correspond to $P=0.99$ (top panel) and $P=0.999$ (bottom panel). Higher-order waveforms only marginally reduce the ambiguity contours.\label{fig4}}
\end{figure}

\subsection{Numerical and systematic effects}
\label{sec:sub:SmallErrors}

At the very smallest scales, delicate implementation-dependent choices can also introduce artificial structure into the
ambiguity function.
We have already extensively described how the choice of reference frequency introduces (coordinate-dependent)
structure. 
Less physically, the sampling rate for the waveforms can produce artificial small-scale structure; to avoid this effect
we sample at a variety of data rates, typically either $8192\unit{Hz}$ (for $P>0.99$) or $16384\unit{Hz}$ (for
$P>0.999$).  
Finally, the ambiguity function can also be impacted by our choice for the starting and ending frequency.   For our
calculations,  we start integrating the waveform and integrate over all power above  $f_{\rm start}=40\unit{Hz}$.
In our experience, this procedure best mimics the real data processing
used in initial LIGO searches.  However, a not-insignificant amount of power is present between $30$ and $40\unit{Hz}$;
if included in the integral, the overlap would differ by  $\Delta P \simeq 10^{-3}$, comparable to some fine-scale
structures of interest.
At the other extreme, we terminate our evolution at the minimum-energy circular orbit (MECO),
where the binary energy ceases to decrease monotonically.



One small, subtle, but important effect is the \emph{nonzero overlap} of the waveforms along the $\pm \hat{z}$
axis\footnote{Such waveforms would have a real overlap of unity, but the complex overlap is expected to be zero. See Appendix \ref{AppA}.}.  For example, for our non-precessing binary, we find that for otherwise identical parameters
\begin{eqnarray}
\text{max}_{t_{\rm c},\psi}|\qmstateproduct{h(\hat{z})}{h(-\hat{z}|t_{\rm c},\psi)}| \simeq 1.7\times 10^{-3}
\end{eqnarray}
Equivalently, in the language of single-detector real overlaps, the sine and cosine chirps are \emph{not precisely
  orthogonal}, for the same orbital phase\footnote{The real overlap is usually performed in conjunction with an
  explicit maximization over time and orbital phase, for a single harmonic.  For that situation, this subtlety does not occur.}.
To a first approximation the $h(\pm\hat{z})$ signals are \emph{basis waveforms}; the waveform along any line of sight is
a superposition of the two.    Because these two signals are not
orthogonal, the ambiguity function generally has fine-scale structure with $\Delta P\simeq 10^{-3}$, associated with the
overlap of these two directions.   For example, on this scale and below, the overlap between two non-precessing
waveforms with just $l=|m|=2$ emission is no longer well-described by Eq.~(\ref{eq:AngularDependence:Leading}).  
Instead, the ambiguity function gains additional fine-scale structure in angle.
%
While extremely useful, for our purposes this result means that on sufficiently small scales  $1-P\lesssim 10^{-3}$, the complex
  overlap will have additional structure compared to ``conventional'' investigations of single-detector,
  optimally-oriented overlaps (i.e., overlaps of two real $h_+$ signals, extracted along $\hat{z}$). 
In particular, this nonzero overlap is partially responsible for the small-scale structure seen in Fig.~\ref{fig3}.

Because of the many subtle interpretation and implementation issues associated with the smallest ambiguity scales, while
we investigate the value of effective fitting to fine scales (e.g., $P>0.999$), for simplicity we emphasize
results for the  scale relevant to most detection events ($P>0.99$).

\section{Results}
\label{sec:Results}
\begin{figure}[!]
\centerline{
 \includegraphics[width=8cm,height=6cm]{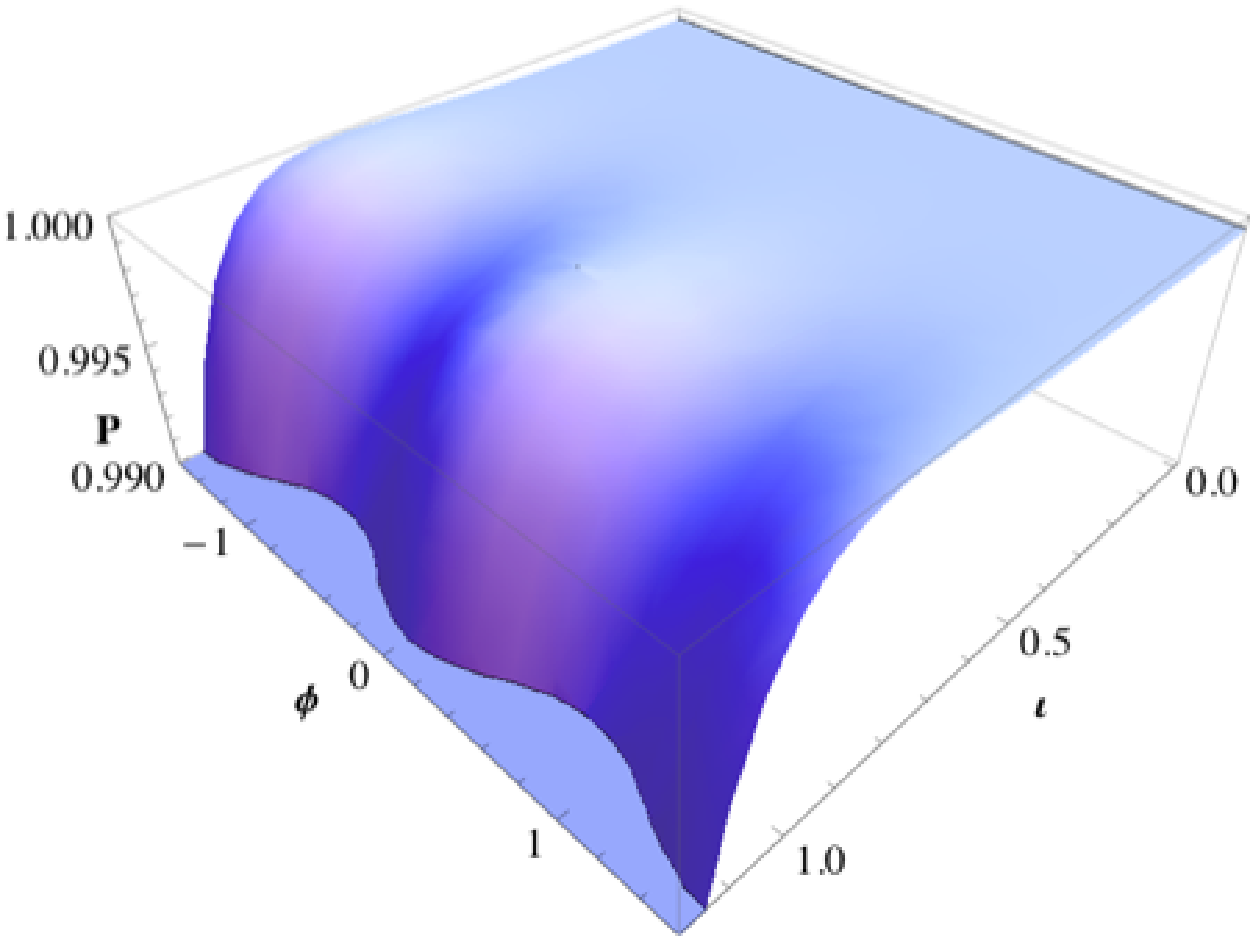}
}
 \centerline{
\includegraphics[width=8cm]{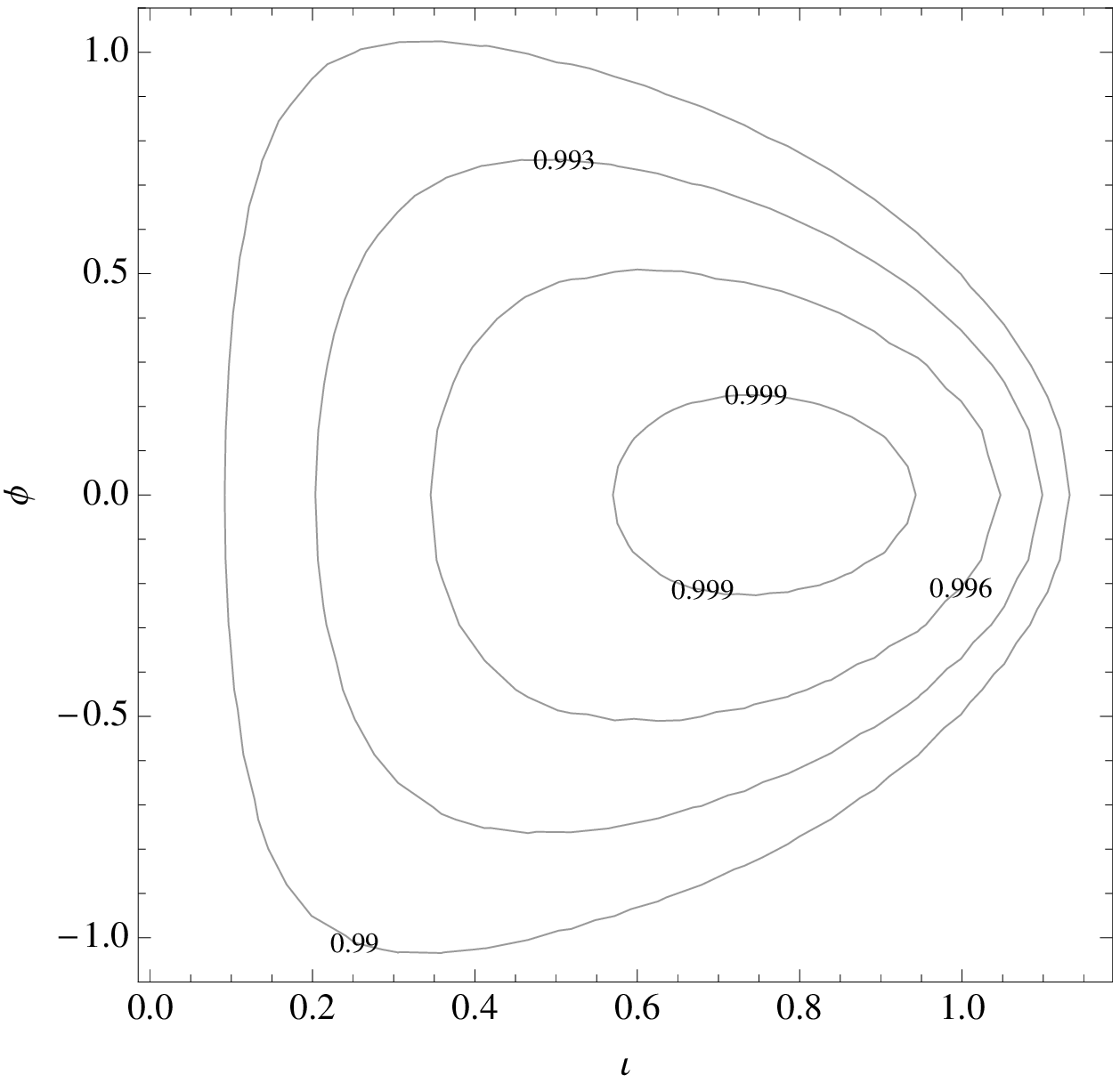}
  }
\caption{{\bf Comparison of the ambiguity surfaces between the leading-order ({\bf top panel}) and higher-order ({\bf bottom panel}) waveforms for the non-spinning binaries.}  ${\bf \delta \lambda}$ is defined by a difference from the fiducial value in the Table~\ref{tab2}. Higher-order waveforms change the ambiguity surface and break degeneracies related to the {\it inclination} and {\it orbital phase} so these parameters can be marginally observable at the scale of $P > 0.99$.\label{fig5}}
\end{figure}

Using a small set of fiducial simulations, we compare non-precessing and precessing signals against their immediate neighbors, mapping out an ambiguity
function in each $n$-dimensional parameter space.  
%

Higher harmonics perturb the ambiguity function by a quantifiable
amount  (i.e., $\delta \Gamma_{ab} \propto v^p$ for $p$ depending on $a,b$ and the harmonic).
As has previously been shown elsewhere, we find that higher harmonics break degeneracies present in non-precessing, 
leading-order signals \cite{Bro07,Lan06,Por08}.   

As described below, we generally find small but significant  \emph{scale-dependent} disagreement with the conventional stationary-phase
Fisher matrix calculation, even in the absence of higher harmonics.    Motivated by Figs.  \ref{fig2} and \ref{fig3},  as well as the Appendix and Figs. \ref{fig:AmbiguityAndReference:1d:Angle} and  \ref{fig:AmbiguityAndReference:2d:McEta}, we
suspect that most scale dependence is introduced by suboptimal coordinates and can be minimized by a better choice of
reference frequency.   
Despite our best attempts to find coordinates well-adapted to the problem, the change in $\Gamma$ going from $P\simeq
0.99$ to $P\simeq 0.999$ for leading-order waveforms is comparable to the change in $\Gamma$ going from leading-order
waveforms to higher harmonics.

\begin{table*}[!]
\begin{tabular}{c | ccc | cc | cc | cc | cc }
  pN order                                            & \multicolumn{7}{|c|}{leading-order}  &      \multicolumn{4}{|c}{higher-order}   \\   
     \hline
 fitting scale                                   &\multicolumn{3}{|c|}{$P > 0.99$}  &\multicolumn{2}{|c|}{$P > 0.999$}
 &\multicolumn{2}{|c|}{$P \sim 1$}  &\multicolumn{2}{|c|}{$P > 0.99$}  &\multicolumn{2}{|c}{$P > 0.999$} \\
     \hline
parameter                                     &         &$M_{\rm c}$ &  $\eta$ &     $M_{\rm c}$   &  $\eta$ &    $M_{\rm c}$   &  $\eta$ &  $M_{\rm c}$   &  $\eta$  & $M_{\rm c}$   &  $\eta$\\
 \hline
\multirow{2}{*}{$(\hat{\Gamma}_{ij})_{\rm eff}$}  & $M_{\rm c}$&  3012$\pm 7$ & -5505$\pm 13$  &     3621$\pm 69$&    -6474$\pm 125$      &2547       &-5314       &3243$\pm 8$      &-6070$\pm 13 $      &3932$\pm 77$  &-7224$\pm 143$  \\
                                                                                       & $\eta$        &                          -&10675$\pm 25$&                                -& 12478$\pm 228$        &     -          &11954       &    -                         &12151$\pm 27$     &  -                        &14367$\pm 270$ \\
  \hline
   \multirow{2}{*}{$c_{ij}$}                                           & $M_{\rm c}$&  1.00 & 0.971                                &1.00    &  0.963     &1.00    &0.963    &   1.00      &0.967    &  1.00      &0.961  \\
                                                                                       & $\eta$            &             -&1.00                              &-            & 1.00        &-    &1.00  &   -    &1.00    &  -    &1.00 \\
  \hline
$ \sigma_i$                                                                  &                          &0.00760&0.00404  &0.00618  & 0.00333&0.00735&0.00339&0.00689&0.00356&0.00578&0.00302 \\
\hline           
   \end{tabular}
 \caption{\label{tab3}{\bf Effective fitting parameters for a non-spinning binary, at different scales.}  The fiducial values
   of parameters are ($M_{\rm c}, \eta$)=(2.994, 0.1077).   Fitting parameters and uncertainties are calculated from a
   least-squares fit, treating the line of sight as fixed.    For the one-dimensional errors $\sigma_i$, we adopt
   $\rho=10$.   Result for $P \sim 1$ is calculated by
   Eq.~(\ref{eq.analyticfisher}) using $h_+$ polarization of the SPA waveforms in $M_{\rm c}, \eta, t_{\rm c}$ and
   $\phi_{\rm ref}$, followed by analytic maximization over $t_{\rm c}$ and $\phi_{\rm ref}$; see the text for details.
   [The effective Fisher matrix corresponds to derivatives of an overlap  maximized over  $t_{\rm c}$ and $\psi$.]
   Systematic differences exist between the SPA waveforms and real overlap used in our numerical calculations and the
   analytic result shown for $P \sim1$.  Nonetheless, all methods largely agree: higher harmonics provide fairly little
   additional information about the chirp mass and mass ratio correlations,with all other parameters fixed; see
   Fig.~\ref{fig4}.   
   For comparison, this table provides fitting parameters for $P$ for two  volumes  ($P>0.99$ and $P>0.999$).  Because
   of the effects described in Fig.~\ref{fig2}, Section \ref{sec:sub:SmallErrors}, and Appendix \ref{ap:ReferenceFrequency}, the two fits disagree.
}
 \end{table*}

\begin{table*}[!]
\begin{tabular}{c | ccccc | 
}
 fitting scale  &\multicolumn{5}{|c|}{$P > 0.99$}                    
 \\
     \hline
parameter                                                                                &         &$M_{\rm c}$ &  $\eta$ & $\iota  $  &  $\phi$   \\
 \hline
\multirow{4}{*}{$(\hat{\Gamma}_{ij})_{\rm eff}$}  & $M_{\rm c}$     &3883$\pm 38$& -7236$\pm 66$ & -0.08350$\pm 0.48$ & 0.08169$\pm 0.23$     \\
                                                                                       & $\eta$             &         -              & 14209  $\pm 138$ & 0.1181$\pm 0.93$ & 1.172$\pm 0.37$          \\
                                                                                       & $\iota  $         &-                        &  -                            & 0.02723$\pm 0.0009$ & -0.001413 $\pm 0.00052$  \\
                                                                                       &  $\phi$        & -                         &-                               &-              & 0.03196  $\pm 0.0003$        \\
  \hline
 \multirow{4}{*}{$c_{ij}$}                                           & $M_{\rm c}$     &1.00& 0.976& -0.00313& -0.271                      \\
                                                                                       & $\eta$             &-        & 1.00& -0.00494& -0.276                  \\
                                                                                       & $\iota  $         &-         &-       & 1.00& 0.0474              \\
                                                                                       &  $\phi$        &-             &-            &-      & 1.00                        \\
                                                                                       \hline
 $ \sigma_i$                                                                  &                        &0.00740& 0.00387& 0.607& 0.583                \\ 
\hline

   \end{tabular}
 \caption{\label{tab4}{\bf Effective fitting parameters for a non-spinning binary, at different scales. Higher
     order}:  Fitting parameters in this table are calculated by least-squares to a quadratic form $P\simeq
   1-(\hat{\Gamma}_{ij})_{\rm eff}\delta \lambda^i \delta \lambda^j/2$.    Measurements of intrinsic ($M_{\rm c},\eta$) and extrinsic
   (line of sight) parameters \emph{separate}: with only one exception $(\hat{\Gamma}_{\eta \phi})_{\rm eff}$, all
   off-diagonal terms coupling the line of sight and intrinsic parameters are  consistent with 0.   We anticipate a
   slightly different choice of  reference frequency will eliminate the small residual correlation that remains.
   Fitting a  general form
   described in Eq.~(\ref{eq:AngularDependence:Higher}) that accounts for the manifestly nonquadratic behavior shown in Fig.~\ref{fig5} leads to comparable results: a separable fit
   (i.e., $G_{c\phi}=0$ and $\Gamma_{aN}\simeq 0$) that performs little better than the quadratic form used above.
  }
 \end{table*}

\begin{table*}[!]
\begin{tabular}{c | ccc | cc | cc | cc | cc }
   pN order                                            & \multicolumn{7}{|c|}{leading-order}  &                                                                                                                 \multicolumn{4}{|c}{higher-order}   \\   
     \hline
 fitting scale                                   &\multicolumn{3}{|c|}{$P > 0.99$}  &\multicolumn{2}{|c|}{$P > 0.999$}  &\multicolumn{2}{|c|}{$P \sim 1$}  &\multicolumn{2}{|c|}{$P > 0.99$}  &\multicolumn{2}{|c}{$P > 0.999$} \\
  \hline
  $P_{\rm max}$                                    &\multicolumn{3}{|c|}{0.999767$\pm 1.09 \times 10^{-5}$}  &\multicolumn{2}{|c|}{0.999884$\pm 0.78 \times 10^{-5}$}  & \multicolumn{2}{|c|}{ n/a}  &\multicolumn{2}{|c|}{0.999775$\pm1.01 \times 10^{-5}$}  &\multicolumn{2}{|c}{0.999888$\pm 0.86 \times 10^{-5}$} \\
     \hline
parameter                                     &         &$M_{\rm c}$ &  $\eta$ &      $M_{\rm c}$   &  $\eta$ &     $M_{\rm c}$   &  $\eta$    & $M_{\rm c}$   &  $\eta$   & $M_{\rm c}$   &  $\eta$\\
 \hline
\multirow{2}{*}{$(\hat{\Gamma}_{ij)_{\rm eff}}$}  &             $M_{\rm c}$&  2899$\pm 8$ & -5295$\pm 14$  &      2980$\pm 62$&    -5325$\pm 111$      &2546       &-5313       &3125$\pm 8$ &-5848$\pm 14$   &3274$\pm 72$ & -6014$\pm 133$  \\
                                                                                                    & $\eta$&-                        &10261$\pm 28$&     -                            & 10276$\pm 207$         &-              &11954      &-                        &11701$\pm 29$  &-                        &11974$\pm 257$  \\
  \hline
   \multirow{2}{*}{$c_{ij}$}                                           & $M_{\rm c}$&  1.00 & 0.971       &1.00& 0.962     &1.00    &0.963    &   1.00      &0.967    &  1.00      &0.960  \\
                                                                                       & $\eta$&             -        &1.00       &-        & 1.00        &-            &1.00  &    -              &1.00    &  -               &1.00 \\
  \hline
$ \sigma_i$                                                                  &                          &0.00776&0.00413  &0.00673  & 0.00363&0.00735&0.00339&0.00704&0.00364&0.00628&0.00328 \\
\hline           
  \end{tabular}
 \caption{{\bf Non-spinning binary fit, using an alternate technique}:   As Table~\ref{tab3}, except the effective
   fitting function allows $P_{\rm max}$ to be a parameter.   In this example, this more generic fit produces  effective Fisher
   matrices that more closely correspond to conventional stationary-phase results for leading-order emission.  
\label{tab3:Alt}
} 
 \end{table*}

\begin{table*}[!]
{\tiny
\begin{tabular}{c | cccc|ccc|ccc|ccc|ccc  }
   pN order                                                                          & \multicolumn{10}{|c}{leading-order}                                                                    & \multicolumn{6}{|c}{higher-order}   \\
     \hline
method                                                                 &\multicolumn{4}{|c|}{iterative}  &\multicolumn{3}{|c|}{simultaneous} &\multicolumn{3}{|c|}{$P \sim 1$}   &\multicolumn{3}{|c|}{iterative}  &\multicolumn{3}{|c}{simultaneous}  \\
     \hline
parameter                                                                   &                         &$M_{\rm c}$ &  $\eta$ & $\chi$ &   $M_{\rm c}$   &  $\eta$ &$\chi$                 &$M_{\rm c}$ &  $\eta$ & $\chi$ &$M_{\rm c}$ &  $\eta$ & $\chi$          &$M_{\rm c}$ &  $\eta$ & $\chi$  \\ 
 \hline
\multirow{3}{*}{$(\hat{\Gamma}_{ij})_{\rm eff}$}  &  $M_{\rm c}$ &3686 &-1652&-1007  &3567$\pm 22$ & -1570$\pm 14$&-975.6$\pm 5.9$&  2837&-2423& -669.6 &4217& -2147&-1177  &4129$\pm 34$&-2083$\pm 25$&-1157$\pm 9.3$  \\
                                                                                              & $\eta$   & -         &1237  &515.5    &-                           &1170$\pm 9$&492.4$\pm 3.5$   &  -       &2357  &612.3 &        -    &1806& 685.6    &           -              &1765$\pm 18$&670.5$\pm 6.4$   \\
			                                                                  & $\chi$&   -        & -           & 283.8   & -                          & -                    &275.5 $\pm 1.6$   &  -      &-           &163.8 &       -    & -       &  340.0      & -                       &      -                  &335.5$\pm 2.6$      \\
 \hline
\multirow{3}{*}{$c_{ij}$}                                           &$M_{\rm c}$&1.00  &-0.947&0.996    &1.00& -0.957&0.997&          1.00 &  -0.981  &0.994                 c&1.00  &-0.929&0.994&               1.00& -0.935&0.995 \\
                                                                                              & $\eta$&-          &1.00&-0.969     & -      & 1.00& -0.974&            -      & 1.00 &   -0.995                   &-         &1.00&-0.958&                   -     & 1.00& -0.962  \\
			                                                                   & $\chi$&-         &-       &1.00       &      -     & -         &1.00&                -  &         - & 1.00                      &   -       &-       &1.00&                        -   & -        &1.00   \\
			                                                           \hline
$ \sigma_i$			                                       &       & 0.0291  &0.0180 & 0.135&             0.0323 &0.0201&0.149&                0.0512&0.0621&0.440                      & 0.0227  &0.0131 & 0.104      & 0.0238 &0.0137&0.108\\
		                                                  
    \end{tabular}
}
 \caption{\label{tab:Fisher:AlignedSpin}
{\bf Effective fitting parameters for an aligned-spin binary.}  As Table~\ref{tab3} for a BH-NS binary with
$\chi=1$.  This binary has one new parameter ($\chi$).  For all parameters shown here, a local quadratic form is a good
approximation to the ambiguity function; this table provides the fitting parameters, treating the line of sight as fixed.
For comparison purposes, this table provides both our standard simultaneous least-squares technique (applied to
$M_{\rm c},\eta,\chi$, with all other parameters fixed) and an iterative technique applied to successive low-order subspaces.  
Table  \ref{tab:Fisher:AlignedSpin:higher} provides the results for a more generic fit including line-of-sight
dependence.
}
 \end{table*}

 \begin{table*}[!]
\begin{tabular}{c | cccccc | 
}
 fitting scale                                                                 &\multicolumn{6}{|c|}{$P > 0.99$}  \\
     \hline
parameter                                                                   &           &  $M_{\rm c}$   &  $\eta$ & $\chi$  & $\iota$  &  $\phi$ \\

 \hline
\multirow{5}{*}{$(\hat{\Gamma}_{ij})_{\rm eff}$}  & $M_{\rm c}$&4388$\pm 54$ & -2256$\pm 40$ & -1230$\pm 15$ & -0.2527$\pm 0.47$ & -0.5585 $\pm 0.38$\\                                                                                 
                                                                                           &  $\eta$&-                                 & 1869$\pm $28& 719.0$\pm $9.8 & 0.3000$\pm $0.31 & 1.122$\pm 0.23$\\
			                                                              & $\chi$&-& -                                                           & 356.2$\pm 4.2$ & 0.1098 $\pm 0.14$& 0.3396$\pm0.11 $\\
			                                                             & $\iota$&-& -& -                                                                                   &0.03279$\pm $0.0009 & -0.001972$\pm 0.0007$\\
			                                                            & $\phi$&-& -& -& -                                                                                                                           & 0.02530$\pm 0.00025$\\
			                                                          \hline
\multirow{5}{*}{$c_{ij}$}                                           & $M_{\rm c}$&1.00& -0.960& 0.997& -0.206& -0.622          \\
                                                                                     & $\eta$&-                & 1.00& -0.976& 0.189& 0.567             \\
			                                                         & $\chi$& - & -                     & 1.00& -0.204   & -0.616            \\
			                                                         & $\iota$&-& -                        & -   & 1.00   & 0.186     \\
			                                                         & $\phi$&-& -& -                              & -             & 1.00           \\
 \hline
$ \sigma_i$			                                       &            & 0.0324& 0.0177& 0.147& 0.566& 0.818             \\
				                                                        \hline                                            
		                                       
    \end{tabular}
 \caption{\label{tab:Fisher:AlignedSpin:higher}
{\bf Effective fitting parameters for an aligned-spin binary. Higher-order}:  As Table~\ref{tab4}, but for a
BH-NS binary with $\chi=1$.
To a first approximation, the Fisher matrix roughly separates the intrinsic and extrinsic parameters.  The weak
covariances shown here are extremely susceptible to small changes in the Fisher matrix coefficients; only a single
preferred value is shown. 
}
 \end{table*}

\begin{table*}[!]
\begin{tabular}{c | cccccccc  }
   pN order                                                                          & \multicolumn{8}{|c}{leading-order}   \\   
     \hline
parameter                                                                   &                  &$M_{\rm c}$   &  $\eta$    & $\chi$    &$\beta$    &$\theta_{\rm NJ}$&$\alpha_{\rm JL}$&$\phi$ \\ 
 \hline
\multirow{7}{*}{$(\hat{\Gamma}_{ij})_{\rm eff}$}  &  $M_{\rm c}$&3234& -281.1& -612.8& 868.8& -1.359& 3.726& -10.18 \\
                                                                                                & $\eta$&-         & 929.9& -3.089& -171.7& 1.252& -6.253& 11.33\\
			                                                                    & $\chi$&-          &-        & 132.0& -177.3& 0.3616& -0.1832& 2.147 \\
			                                                                   &$\beta$& -        &-         &-          & 295.2& 0.06336& 1.674& -6.945\\
			                                                 &$\theta_{\rm NJ}$&-           &-        &-          &-         & 0.7774& -0.003074& 0.006970\\
			                                                 &$\alpha_{\rm JL}$&-          &-          &-          &-        &-           & 0.3349& -0.1021\\
			                                                                     &$\phi$&-         &-           &-           &-       &-          &-              & 0.3829\\
 \hline
 \multirow{6}{*}{$c_{ij}$}                                         &   $M_{\rm c}$&1.00 & 0.407 & 0.730 & 0.144 & -0.0729 & -0.109& -  \\
                                                                                                 & $\eta$&-       & 1.00 & 0.742 & 0.729 & -0.183 & 0.109& -\\
         		                                                                    & $\chi$&-&-                         & 1.00 & 0.752 & -0.180 & -0.133& - \\
         		                                                                   &$\beta$& -&-&-                                & 1.00 & -0.191 & -0.106& -\\
         		                                                 &$\theta_{\rm NJ}$&-&-&-&-                                            & 1.00 & 0.0126& -\\
         		                                                 &$\alpha_{\rm JL}$&-&-&-&-&-                                                  & 1.00& -\\
         		                                                           \hline
$ \sigma_i$			                                       &          &0.00592& 0.00573& 0.0433& 0.0212& 0.116& 0.187& - \\
			                                                         
			                                                        \hline
			                                                         \hline

   pN order                                                                          & \multicolumn{8}{|c}{higher-order}   \\   
     \hline
parameter                                                                   &                  &$M_{\rm c}$   &  $\eta$    & $\chi$    &$\beta$    &$\theta_{\rm NJ}$&$\alpha_{\rm JL}$&$\phi$ \\ 
 \hline
\multirow{7}{*}{$(\hat{\Gamma}_{ij})_{\rm eff}$}  &  $M_{\rm c}$&3401& -408.2& -644.3& 911.9& -1.745& 4.515& -11.89  \\
                                                                                                & $\eta$&-       & 1055& 7.249& -208.3& 1.495& -7.229& 12.58\\
			                                                                    & $\chi$&-&-                  & 139.3& -186.3& 0.5298& -0.1706& 2.471 \\
			                                                                   &$\beta$& -&-&-                        & 307.9& 0.02350& 1.998& -7.544\\
			                                                 &$\theta_{\rm NJ}$&-&-&-&-                                   & 0.8305& 0.003262& 0.006150\\
			                                                 &$\alpha_{\rm JL}$&-&-&-&-&-                                               & 0.3510& -0.1029\\
			                                                                     &$\phi$&-&-&-&-&-&-                                                            & 0.3951\\
 \hline
\multirow{6}{*}{$c_{ij}$}                                           & $M_{\rm c}$&1.00 & 0.543 & 0.778 & 0.309 & -0.154 & -0.177& -  \\
                                                                                                 & $\eta$&-      & 1.00 & 0.815 & 0.786 & -0.265 & -0.00409& -\\
 			                                                                    & $\chi$&-&-               & 1.00 & 0.815 & -0.270 & -0.230& - \\
 			                                                                   &$\beta$& -&-&-                    & 1.00 & -0.271 & -0.204& -\\
 			                                                 &$\theta_{\rm NJ}$&-&-&-&-                            & 1.00 & 0.0381& -\\
 			                                                 &$\alpha_{\rm JL}$&-&-&-&-&-                                    & 1.00& -\\
 			                                                           \hline
$ \sigma_i$			                                       &          &0.00626& 0.00626& 0.0495& 0.0237& 0.115& 0.188& - \\

    \end{tabular}
 \caption{\label{tab:Fisher:Precessing:Iterative}\textbf{Effective fit for precessing binary: Iterative method}:  The effective Fisher matrix
   needed to reproduce the calculated ambiguity function, derived from an iterative fit; see Table~\ref{tab:Fisher:Precessing} for
   a full 
   $7$-dimensional fit.    Both of these matricies have comparable lists of eigenvalues
   $(3625,908,54,4.9,0.74,0.29, \simeq 0)$  and $(3844,999,57, 4.1,0.75, 0.28, \simeq 0)$ [bottom] and
with the nearly zero eigenvalue roughly corresponding to the $\phi$ direction.  
Since the eigenvectors span
several orders of magnitude, the correlation coefficients of this poorly-conditioned matrix are extraordinarily sensitive
to small changes in the coefficients and have not been provided.     Unlike the non-precessing case, the geometric and intrinsic
parameters do not completely separate, even for these well-adapted coordinates.   
The one-parameter uncertainties $\sigma_i$ shown are calculated  by omitting the (nearly unmeasurable) $\phi$ direction
from the fit.
}
 \end{table*}

\begin{table*}[!]
\begin{tabular}{c | cccccccc  }
  pN order                                                                          & \multicolumn{8}{|c}{leading-order}   \\   
     \hline
parameter                                                                   &                  &$M_{\rm c}$   &  $\eta$    & $\chi$    &$\beta$    &$\theta_{\rm NJ}$&$\alpha_{\rm JL}$&$\phi$ \\ 
 \hline
\multirow{7}{*}{$(\hat{\Gamma}_{ij})_{\rm eff}$}  &  $M_{\rm c}$&3279$\pm 12$ & -291.4$\pm 20$ & -618.8$\pm 2.7$ & 878.7$\pm 3.4$ & -2.168$\pm 0.68$ & 3.982$\pm 0.38$              & -11.58$\pm 0.32$ \\
                                                                                                & $\eta$&-                          & 922.8 $\pm 5.3$& -5.187$\pm 4.5$ & -175.5$\pm 4.7$ & 1.648$\pm 0.34$ & -6.132 $\pm 0.17$           & 11.80$\pm 0.12$\\
			                                                                    & $\chi$&-                         & -                              & 132.3$\pm 0.52$ & -177.9$\pm 0.69$ & 0.4429$\pm 0.13$ & -0.2340$\pm 0.079$ & 2.321$\pm 0.063$\\
			                                                                   &$\beta$& -                       & -                              & -                              & 295.1$\pm 1.5$& 0.1488$\pm 0.20$ & 1.674 $\pm 0.11$             & -7.263 $\pm 0.058$\\
			                                                 &$\theta_{\rm NJ}$&-                        & -                               & -                            & -                             & 0.8038$\pm 0.0046$ & -0.004911 $\pm 0.0063$& 0.005865$\pm 0.0069$\\
			                                                 &$\alpha_{\rm JL}$&-                       &-                               &-                              &-                              &-                                   & 0.3299 $\pm 0.0018 $     & -0.1100$\pm 0.0035$\\
			                                                                     &$\phi$&-                       &-                                &-                              &-                              &-                                   &-                                           & 0.3850$\pm 0.0020$\\
 \hline
\multirow{6}{*}{$c_{ij}$}                                         &   $M_{\rm c}$&1.00 & 0.477 & 0.747 & 0.238 & -0.0961 & -0.104& -  \\
                                                                                                & $\eta$&-        & 1.00 & 0.807 & 0.792 & -0.265 & 0.0961& -\\
			                                                                    & $\chi$&-&-                & 1.00 & 0.801 & -0.251 & -0.108& - \\
			                                                                   &$\beta$&-&-&-                     & 1.00 & -0.284 & -0.0817& -\\
			                                                 &$\theta_{\rm NJ}$&-&-&-&-                            & 1.00 & 0.0119& -\\
			                                                 &$\alpha_{\rm JL}$&-&-&-&-&-                                    & 1.00& -\\
			                                                           \hline
$ \sigma_i$			                                       &          &0.00622&0.00661&0.0495&0.0241&0.117&0.187& - \\
			                                                         
			                                                        \hline
			                                                         \hline

  pN order                                                                          & \multicolumn{8}{|c}{higher-order}   \\   
     \hline
parameter                                                                   &                  &$M_{\rm c}$   &  $\eta$    & $\chi$    &$\beta$    &$\theta_{\rm NJ}$&$\alpha_{\rm JL}$&$\phi$ \\ 
 \hline
\multirow{7}{*}{$(\hat{\Gamma}_{ij})_{\rm eff}$}  &  $M_{\rm c}$&3403$\pm 13$ & -420.2 $\pm $ 21& -640.4$\pm  2.4$ & 916.4 $\pm 3.6$& -2.845$\pm 0.74$ & 5.016 $\pm 0.40$              & -12.35$\pm 0.31$  \\
                                                                                                & $\eta$&-                      & 1057$\pm 6.4$      & 18.55 $\pm 4.9$& -212.3$\pm 5.3$ & 1.533 $\pm 0.38$& -7.488 $\pm 0.20$                & 12.73$\pm 0.13$\\
			                                                                    & $\chi$&-&-                                                        & 137.2 $\pm 0.55$& -185.2 $\pm 0.73$& 0.6655 $\pm 0.14$& -0.2104$\pm 0.090$       & 2.472$\pm 0.061$ \\
			                                                                   &$\beta$&-&-&-                                                                                      & 308.5$\pm 1.2$ & -0.01406$\pm 0.22$ & 2.022$\pm 0.12$            & -7.594$\pm 0.060$\\
			                                                 &$\theta_{\rm NJ}$&-&-&-&-                                                                                                                    & 0.8481 $\pm 0.0048$& 0.003203 $\pm 0.0067$& 0.01161$\pm 0.0068$\\
			                                                 &$\alpha_{\rm JL}$&-&-&-&-&-                                                                                                                                                        & 0.3456 $\pm 0.0018$    & -0.1035$\pm 0.0038$ \\
			                                                                     &$\phi$&-&-&-&-&-&-                                                                                                                                                                                                   & 0.3911$\pm 0.0020$\\
 \hline
\multirow{6}{*}{$c_{ij}$}                                           & $M_{\rm c}$&1.0 & 0.391 & 0.726 & 0.127 & -0.0705 & -0.193& -  \\
                                                                                                & $\eta$&-     & 1.00 & 0.718 & 0.712 & -0.226 & 0.0670& -\\
			                                                                    & $\chi$&-&-             & 1.00 & 0.744 & -0.224 & -0.234& - \\
			                                                                   &$\beta$&-&-&-                   & 1.00 & -0.252 & -0.170& -\\
			                                                 &$\theta_{\rm NJ}$&-&-&-&-                            & 1.00 & 0.0224& -\\
			                                                 &$\alpha_{\rm JL}$&-&-&-&-&-                                       & 1.00& -\\
			                                                           \hline
$ \sigma_i$			                                       &          &0.00584&0.00529&0.0421&0.0210&0.113&0.191& - \\

    \end{tabular}
 \caption{\label{tab:Fisher:Precessing}\textbf{Effective fit for precessing binary: Simultaneous method}
The effective Fisher matrix
   needed to reproduce the calculated ambiguity function, derived from a full $7$-dimensional least squares; see Table~\ref{tab:Fisher:Precessing:Iterative}  for
   an alternative iterative fit.    Both methods produce comparable results, with comparable eigenvalue distributions.    The difference between the leading-order result and a model
   including higher harmonics is small, though usually significantly in excess of our fitting parameter error (e.g.,
   several standard deviations) and of the systematic differences between the two fitting methods.  As expected, higher
   harmonics lead to \emph{smaller} parameter correlations
}
 \end{table*}

\subsection{Zero spin}

For a system without spin, higher harmonics principally provide information about the line of sight.    For clarity, we
will first discuss the most immediately relevant scale ($P\gtrsim 0.99$).   Comparing the solid (without higher harmonics)
and dotted (with higher harmonics) curves on the top panel of  Fig.~\ref{fig4}, we immediately see that higher
harmonics provide little new information about intrinsic parameters, all other things being equal.   
Equivalently, looking at Table~\ref{tab3}, the effective Fisher matrix $\hat{\Gamma}$ on the two-dimensional parameters
$M_{\rm c},\eta$ without and with higher harmonics are similar to each other, as well as to a standard Fisher matrix computed using stationary phase approximation waveforms
(labeled as $P\sim 1$). 
By contrast, as illustrated by the dramatic difference between the top and bottom panel in Fig.~\ref{fig5},  higher harmonics produce a dramatic qualitative change in how well the line of sight $\hat{n}$ can be
measured.  
%

Both with and without higher harmonics [Fig.~\ref{fig5}], the line of sight is very difficult to measure, particularly at the expected
relevant scale $P\simeq 0.99$ (i.e., SNR of around $10$).   In both cases, the ambiguity function has a broad, extended,
 asymmetric extremum.   In the absence of higher harmonics, the ambiguity function \emph{cannot} be usefully described by a locally quadratic approximation, even a
effective one.   
Nonetheless, by understanding the expected  dependence on angle (and by adopting coordinates in band), we can propose a
physically-well-motivated fitting function, both for the purely angular dependence and for the correlations between line
of sight and other parameters [Eq.~(\ref{eq:AngularDependence:Higher})].  
This fitting function works extremely
well when higher harmonics are neglected.   When higher harmonics are included, a quadratic approximation sufficies, as the local extremum is much narrower.   
In the latter case, Table~\ref{tab4} provides the fitting parameters needed to reconstruct the full multidimensional
fit. 
In fact, our well-chosen reference frequency produces a \emph{nearly separable fit}, with zero off-diagonal terms
[e.g., $\Gamma_{aN}\simeq 0,G_{\phi c}\simeq 0$ in Eq.~(\ref{eq:AngularDependence:Higher})]. 
We will return to this simple structure frequently below.

%
Several effects besides higher harmonics also introduce fine-scale structure into the ambiguity function. 
As demonstrated in Fig.~\ref{fig2}, the choice of reference frequency can introduce strong, scale-dependent features
into the ambiguity function.   We chose a reference frequency at $100 {\rm Hz}$ to reduce its effect, but have not
eliminated it completely.   
The nonzero overlap between the $(l,m)=(2,\pm 2)$ modes is another such effect.
Hence, we are not surprised that our effective fitting parameters change  as we reduce the range of
$P$ used in the fit, even in the absence of harmonics; see  Table~\ref{tab3}.
%

The effect of scale dependence is fairly mild: the eigendirections for $P>0.99$ and $P>0.999$ agree, only the eigenvalue
scale changes. 
%
For comparison, we also considered an alternate fitting technique that allowed the single best fit point to have
$P_{\rm max}\ne 1$; see Fig.~\ref{fig3}, and Table~\ref{tab3:Alt}.
While this method leads to aesthetically pleasing results similar to analytic calculations,  this fit systematically
underestimates $P$ near the maximum-likelihood point and does not completely eliminate the trend towards different
fitting parameters on the smallest scales.  We henceforth adopt $P_{\rm max}=1$.

To facilitate approximate comparisons with prior work, Tables~\ref{tab3} and \ref{tab4} provide one-dimensional standard deviations
$\sigma$ and correlation coefficients $c_{ij}$.  Unless otherwise stated, these quantities are derived \emph{solely from
  the Fisher matrix fits from that same table}.   For example, in
Table~\ref{tab3}, the ``measurement errors'' $\sigma_i$ follow from inverting the $2\times 2$ matrix shown, while in
Table~\ref{tab4} they follow  from inverting a full $4\times 4$ matrix.

\subsection{Aligned spin}

\begin{figure}
\includegraphics[width=\columnwidth]{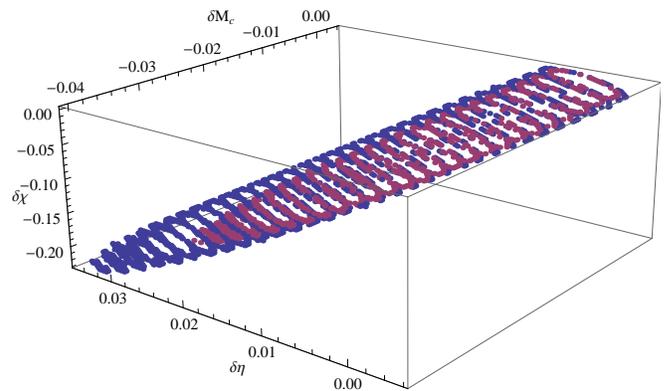}
\caption{\label{fig:AlignedSpin:ThreeDErrorEllipsoids} \textbf{Ambiguity function for aligned-spin}: Points in
  $M_{\rm c},\eta,\chi$ with $P\in[0.99, 0.991]$, shown as blue (leading-order) and dark red (with higher harmonics).  The two long,
  narrow surfaces (ambiguity ellipsoids) these points cover illustrates gravitational wave measurements can constrain one combination of
  $M_{\rm c},\eta,\chi$ tightly (e.g., $M_{\rm c}$); one less so (e.g., $\eta$); and one almost not at all.
  The two surfaces nearly agree,    with the most significant change being a slight reduction in the least-well-determined
  direction.  The close agreement between these two surfaces shows higher harmonics provide fairly little additional information to
  break this degeneracy. 
}
\end{figure}

%
Repeating our effective Fisher matrix calculation for an aligned-spin BH-NS binary leads to results qualitatively
similar to the zero-spin case.   
As previously, higher harmonics  provide little additional information about intrinsic parameters, here $M_{\rm c},\eta,\chi$;
see Fig.~\ref{fig:AlignedSpin:ThreeDErrorEllipsoids}.   For example, looking at data along a fixed line of sight,  Table~\ref{tab:Fisher:AlignedSpin} shows that, on
a component-by-component basis ($\Gamma$) and overall ($\sigma_i$), the two ambiguity functions with and without higher
harmonics resemble one another.   
More directly,  Fig.~\ref{fig:AlignedSpin:ThreeDErrorEllipsoids} shows three-dimensional contours of nearly constant
$P$ as a function of $M_{\rm c},\eta,\chi$ for both leading-order emission (blue) and higher harmonics (red).  While higher
harmonics clearly do provide more information about intrinsic parameters -- the red surface is nested inside the blue --
the addition of higher harmonics only marginally improves our ability to measure the least-well-constrained combination
of $M_{\rm c},\eta,\chi$.  
Precisely as in the non-spinning case, however, higher
harmonics provide more information about the line of sight. 
Despite our line of sight providing some sensitivity to a symmetry-breaking $(2,1)$ harmonic, a waveform with harmonics
encodes roughly similar information as a waveform without harmonics. 
%
%
Tables~\ref{tab:Fisher:AlignedSpin} and \ref{tab:Fisher:AlignedSpin:higher} provide the effective fitting
parameters
we used to reproduce their ambiguity function.  
As in the non-spinning case, we find an  \emph{approximately} separable fit, though less so than before [see $c_{ij}$ in
  Table~\ref{tab:Fisher:AlignedSpin:higher}].
Even allowing for weak correlations between intrinsic parameters and the line of sight, the overall parameter
covariances $\sigma_i$ with harmonics [Table~\ref{tab:Fisher:AlignedSpin:higher}] are nearly unchanged from a model with
only leading-order emission [Table~\ref{tab:Fisher:AlignedSpin}].\footnote{For leading-order emission,
  measurements of orientation and intrinsic parameters separate completely.  The intrinsic parameter uncertainties computed in   Table
  \ref{tab:Fisher:AlignedSpin} are therefore identical to the uncertainties for the corresponding parameters
  derived from a full $5\times 5$ Fisher matrix.}

The aligned-spin results at $\chi=1$ cannot be easily compared with the corresponding zero-spin results ($\chi=0$).  On  the one hand,
the component-by-component  Fisher matrix coefficients like $\Gamma_{\eta \eta}$ will differ significantly, as the
waveform phasing changes as a function of $\chi$ and hence so does $\left<(\partial_\eta \Psi)^2\right>$.   On the other
hand, the aligned-spin results allow a \emph{new parameter} (spin) that was  treated as fixed for the $\chi=0$ case, with nontrivial
coupling to the other intrinsic parameters.  The one-parameter uncertainties $\sigma_i$ are \emph{dramatically}
increased by including this previously-neglected systematic effect.

Both at leading and higher-order, our effective fit to the ambiguity function is complicated by the wide range of
scales in $\Gamma$, even for fixed line of sight.   As is well-known from previous Fisher matrix calculations with
aligned-spins \cite{Poi95,Cut94},  the ambiguity function in $M_{\rm c},\eta,\chi$ has strong correlations, producing a narrow and
extended extremum.     For the specific example described by our  effective Fisher matrix, the $M_{\rm c},\eta,\chi$ submatrix has eigenvalues $\simeq
4200, 400, 30$, describing a strong  hierarchy of scales.    
For our purposes,    Fig.~\ref{fig:AlignedSpin:ThreeDErrorEllipsoids} demonstrates our 
 fiducial aligned-spin binary cannot be distinguished from binaries with spin $\chi  \gtrsim 0.8$: for each $\chi$ in
 this range, suitable combinations of $M_{\rm c},\eta$
 exist with high overlap.    
%
For these extremely extended ambiguity ellipsoids, a fit that reproduces $P(\lambda_0,\lambda)$ over the full range in $\chi$ \emph{might} require a more
generic functional form  than the one adopted so far: a  quadratic with constant coefficients \emph{in the neighborhood
  of the fiducial binary}.
%
Effectively speaking, however, these additional degrees of freedom add little information with considerable
expense.  
%
We will explore more complicated effective dependence in a subsequent publication.


As in the zero-spin case, we find significant differences on the smallest scales in $P$, in a fashion that depends on the
reference frequency.
Given the number of dimensions, complex functional form, sensitivity to numerical implementation like the sampling rate,
and less immediate observational relevance, we defer a detailed discussion of fine-scale effects to a subsequent paper.


\subsection{Precessing spin: Case 1}

For the first of our two fiducial precessing binaries, we find higher harmonics provide little added information beyond the constraints
produced in the non-precessing case.   This unfortunate but expected result can be seen, for example, from the one-dimensional covariances in Table
\ref{tab:Fisher:Precessing}; from  the effective Fisher matrix coefficients $\hat{\Gamma}_{ij}$; or from their
comparable sequences of eigenvalues.  
%
That said, even in the absence of higher harmonics, the ambiguity function for a \emph{precessing} binary has  simpler structure than the
non-precessing result, with  reduced correlations among the ``intrinsic'' parameters; a somewhat less
extreme hierarchy of scales (i.e., eigenvalues)\footnote{We can always rescale our eigenvalues by rescaling our coordinate units.  However, in these units all parameters have a prior range of order unity, and physical meaning.  As a result, our eigenvalues also have meaning: the coordinate combinations corresponding to the smallest  eigenvalues have minimal impact on the overlap and can be ignored.}; and roughly speaking a more quadratic ambiguity function.  

In fact, for a precessing binary the previous clear separation between ``intrinsic'' and ``geometric'' parameters breaks down.   As each instant the opening angle $\beta$ of the precession cone of $L$ around $J$
depends on the relative magnitude of $L$ and $S=J-L$, as well as on their (nearly conserved) misalignment angle $\hat{L}\cdot
\hat{S}$.   The magnitudes of $L$ and $S$ are essentially \emph{intrinsic} parameters, characterizing the binary masses
and BH spin; therefore, we expect the precession cone opening angle  $\beta$ to be intimately correlated with the
intrinsic parameters.  
At the same time, the precession cone opening angle must be intimately connected to the ``geometric'' parameters that define the
orientation of the binary at the reference frequency: $\theta_{NJ},\alpha_{JL},\phi$.   Specifically, the orientation of the precession of $L$ relative to the line of sight is characterized by the two
angles $\theta_{NJ}$  (setting the orientation of $J$)\footnote{The other angle needed to specify the orientation of $J$
  is equivalent to a rotation around the line of sight, i.e. the polarization angle.  As the complex overlap maximizes over this angle, it is
  explicitly removed as a parameter.} and $\alpha_{JL}$ (fixing the orientation of $L$ along the precession cone at the
reference frequency).   The orbital phase $\phi$ at the reference frequency fixes the binary's geometry in band.  
This cross-coupling between intrinsic and extrinsic parameters has  quantitative consequences for the Fisher matrix.
Roughly speaking, the two new eigenvalues
introduced into $\Gamma$ by allowing precession 
(here with values $\simeq 5, 0.7$, associated with the $\beta$ and $\alpha_{JL}$ parameters) can be expected to  lie \emph{between} the very large
($3000,900,100$) and very small ($\simeq 0.3,\simeq 0$) eigenvalues associated with the manifestly intrinsic ($M_{\rm c},\eta,\chi$)
and extrinsic ($\theta_{NJ},\phi$) parameters.

For non-precessing binaries,  the choice of  a $100$ Hz reference frequency nearly separated intrinsic and extrinsic
parameters.  A suitable choice of reference frequency may yet further reduce the off-diagonal terms in our
effective Fisher matrix. 
For the present coordinates, however, we cannot cleanly decompose parameters into ``intrinsic'' and ``geometric''
parameters.  Table~\ref{tab:Fisher:Precessing} shows correlation coefficients $c_{ij}$ calculated by omitting the
(nearly unmeasurable) $\phi$ coordiante in $\Gamma$; no obvious block-diagonal form occurs.

%
%

\subsection{Precessing spin: Case 2}
\begin{figure}
\includegraphics[width=\columnwidth]{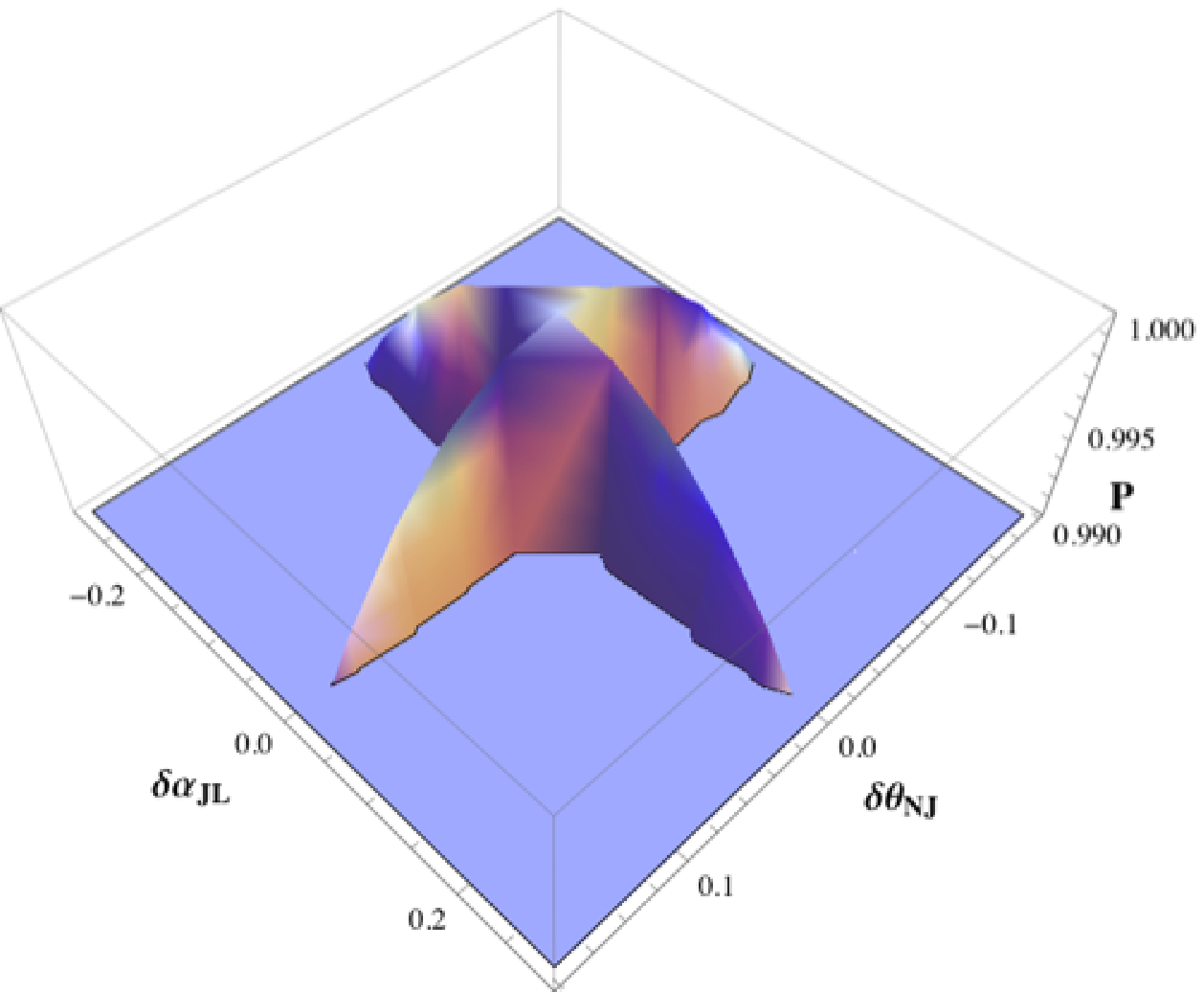}
\includegraphics[width=\columnwidth]{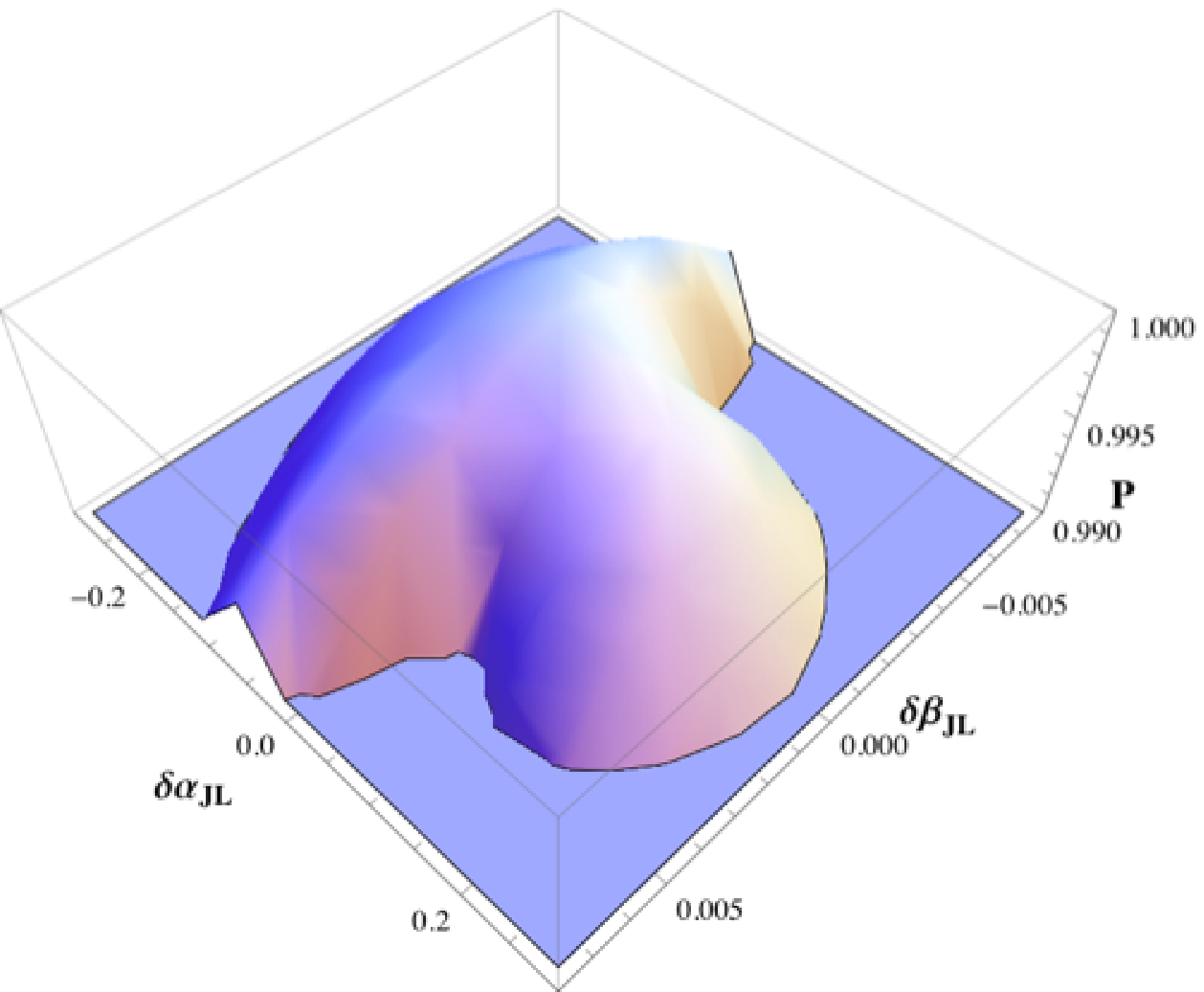}
\caption{\label{fig:Precessing:Case2:HorribleNonquadratic}\textbf{Symmetry can produce highly nonquadratic behavior}: Plot of the ambiguity function when changing only
  $\theta_{NJ}$ and $\alpha_{JL}$ (top panel) and $\beta_{JL},\alpha_{JL}$ (bottom panel), for the high-symmetry binary
  parameter set ``case 2''.}
\end{figure}

By contrast to the relatively simple ambiguity functions seen so far, our second set of binary parameters produces a
significantly more complicated ambiguity function, particularly in geometric parameters.
For example, Fig.~\ref{fig:Precessing:Case2:HorribleNonquadratic} shows the ambiguity function versus
$\theta_{NJ},\alpha_{JL}$ and $(\beta_{JL},\alpha_{JL})$ all other parameters fixed, for ``case 2''. As in case 1, 
we have a highly symmetric binary starting with $L,J$, and the line of sight in the same plane at our reference frequency. However, in this case we start with $L$ parallel to the line of sight, rather than perpendicular to it.  The ambiguity function shows extreme sensitivity
to the initial conditions and highly nonquadratic behavior.   
These differences occur despite the considerable similarity between case 1 and case 2: the two are, to an excellent
approximation, the same configurations, just slightly offset in time. 
By contrast, the change in ambiguity versus $M_{\rm c},\eta,\chi$ is well-described by a quadratic form. 
This extreme scenario demonstrates that even an \emph{effective} Fisher matrix has limits: sometimes, a more
generic functional form including higher-order correlations must be used on relevant scales.
That said, highly nonquadratic behavior only occurred for a high-symmetry binary.  We expect typical binary initial
conditions will produce nearly quadratic ambiguity functions.

\section{Conclusions}

In this paper we have used case studies of a coherent, two-detector  ambiguity function $P(\lambda_0,\lambda)$ to estimate how much and what kind of information higher harmonics provide
about BH-NS binaries.  
Given the high dimension of and severe degeneracies that plague the problem, we perform a tractable, idealized calculation
instead of the straightforward but less easily understood explicit source in a real multi-detector network.   Specifically, we place a
single source directly overhead an idealized detector pair, equally sensitive to both polarizations. 
%
For  all binaries, we find that higher harmonics provide little additional information about the binary's intrinsic
parameters.   Instead, at best they provide information about the orientation of the source relative to the line of
sight (e.g., $\iota,\phi$ for non-precessing binaries).
Notably, higher harmonics make it easier to exclude a non-precessing binary with $\hat{L} \simeq  \pm \hat{n}$.
%

When possible, we estimate the two-point function $P(\lambda_0,\lambda)$ on scales of interest ($1-P<1/\rho^2$) by a locally quadratic function
$1-\Gamma_{\alpha\beta}\delta \lambda^{\alpha}\delta \lambda^{\beta}/2$, whose coefficients we denote the ``effective Fisher
matrix.''   
Using coordinates $\lambda$ adapted to the most sensitive frequencies of our network, we show that to an excellent approximation
the Fisher matrix for non-precessing binaries separates into intrinsic (masses and spins) and extrinsic (geometric)
parameters.
For precessing binaries, the angles describing the precession cone are intimately connected with the intrinsic masses and spins as well as the geometric parameters. Therefore, they correlate with both intrinsic and extrinsic parameters and entangle the two previously distinct sets of parameters.
%
In future work we will compare our results to detailed posterior parameter distributions, as computed by Markov-chain
Monte Carlo simulations.  
Our results suggest suitable coordinates will substantially simplify the interpretation of these posteriors.  
%

In addition to our main results, this paper also provides technical suggestions of broad interest to the data analysis
community.
First, rather than performing the conventional (single-polarization, real) overlap maximized over time and orbital phase,
we use a coherent two-polarization overlap  maximized over time and polarization \cite{Osh11}.  This overlap is far more
discriminating than the single-polarization result and, lacking a maximum over the polar emission direction $\phi$, is
well adapted to asymmetric situations such as precession or the presence of higher harmonics.  
%
Second, we illustrate the importance of choosing coordinates adapted to the network's sensitive band and to the binary's in-band geometry.  We have provided concrete examples of the consequences of poor choices for coordinate conventions, which can lead to pathological behavior in the complex overlap and ambiguity function.
%
Finally, we recommend adopting an effective Fisher matrix, derived by fitting the ambiguity function on the scale at which 
variations in the parameters could be plausibly detected.   We have shown that the conventional approach to the 
Fisher matrix can be sensitive to unobservable fine-scale structure which can give misleading results.    
By contrast, simply by computing the ambiguity function over the area of interest, then fitting, one can explicitly verify whether a quadratic approximation even applies, as well as assess its error.
\section*{Acknowledgements}
HSC, CK and CHL are supported in part by National Research Foundation Grant funded by the Korean Government 
(NRF-2011-220-C00029) and the Global Science experimental Data hub Center (GSDC) at KISTI.
HSC and CHL are supported in part by the BAERI Nuclear R \& D program (M20808740002) of MEST/KOSEF.
CK is also supported by the Research Corporation for Scientific Advancement 
and by a WVEPSCoR Research Challenge Grant.
ROS is supported by NSF award PHY-0970074, the Bradley Program Fellowship, and the UWM Research
Growth Initiative. 
EO is supported by NSF award PHY-0970074.

\appendix
\section{Comparison between the Real and Complex overlaps} \label{AppA}
One detector can not be sensitive to both polarizations, as it measures a single (real) strain variable $h$.  The
conventional definition of the overlap [Eq.~(\ref{eq.conventionaloverlap})] therefore provides an inner product for a single real
data sequence.

\begin{figure}[!]
\centerline{
\includegraphics[width=\columnwidth]{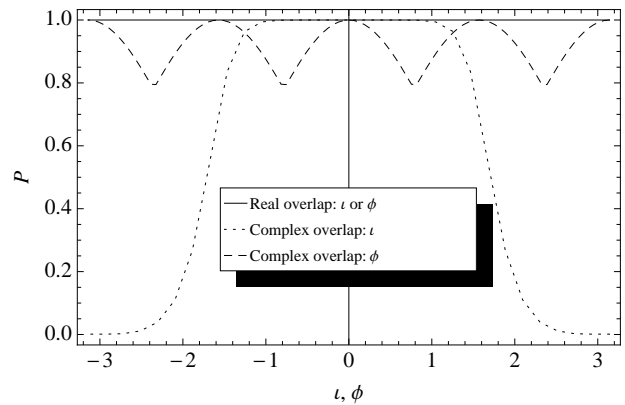}
 }
\caption{{\bf Comparison of the ambiguity functions between a real and complex overlaps for the extrinsic parameters.} Dotted line is calculated by changing $\iota$ where the fiducial value is 0 and other parameters are the same as in Table~\ref{tab1}. Dashed line is calculated by changing $\phi$ where $\iota=\pi/2$ and other parameters are the same as in Table~\ref{tab1}. The ambiguity surface for the real overlap is flat.
\label{fig.realversuscomplex1}} 
\end{figure}

\begin{figure}[!]
 \centerline{
\includegraphics[width=\columnwidth]{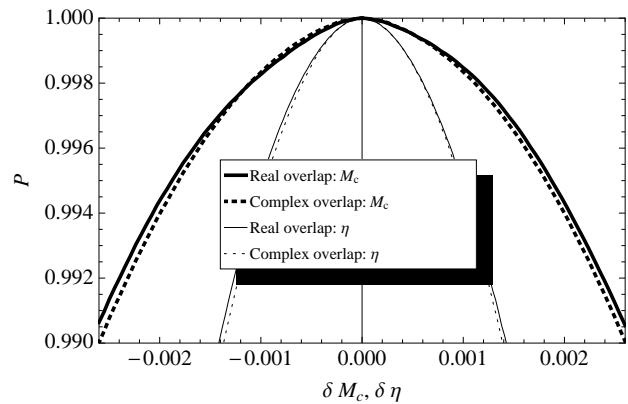}
 }
\caption{{\bf Comparison of the ambiguity functions between a real and complex overlaps for the intrinsic parameters.} Parameter values are summarized in Table~\ref{tab1}. Thick lines are calculated by changing $M_{\rm c}$, others by changing $\eta$.
\label{fig.realversuscomplex2}} 
\end{figure}

The complex overlap [Eq.~(\ref{eq.complexoverlap})], however, uses information about both polarizations \cite{Osh12}.
To illustrate the differences between these two diagnostics, Figs. \ref{fig.realversuscomplex1} and \ref{fig.realversuscomplex2}
show the overlaps $\qmstateproduct{h_0}{h}$ between our fiducial non-spinning waveform  and an identical binary,
except for one parameter.  In both cases, these overlaps are maximized over event time and polarization (for the
complex overlap) or orbital phase (for the real overlap). 

First and foremost, Fig.~\ref{fig.realversuscomplex1} shows that the \emph{real} overlap does not change as 
inclination ($\iota$) or the orbital phase ($\phi$) are varied.  In other words, with one detector, we cannot identify the
inclination, which measures the relative amplitude of $h_+$ to $h_\times$.  Also, by maximizing over $\phi$, we
lose information about it.  The conventional single-detector overlap therefore provides no
information about how well we can measure these parameters in a network sensitive to two polarizations. 

By  contrast, because the complex overlap explicitly uses two polarizations, it can identify the inclination; see the dotted
line in Fig.~\ref{fig.realversuscomplex1}.  
 For example, because  it can distinguish between left- and right-handed
sources, the overlap between antipodal directions (i.e., $\iota=0, \iota=\pi/2$) is zero. 
Moreover, if the source orientation is different from $0$, then the waveform carries information about the orbital
phase, which the complex overlap easily identifies.    
As a concrete example, if $\iota=\pi/2$ for the fiducial signal, the orbital phase $\phi$ is measurable at high
SNR. When $\iota=\pi/4$, $\phi$ is unmeasurable for the leading-order waveforms (for the analytic formula, see Eq.~(B1)
in \cite{Osh12}) but marginally  measurable for the higher-order waveforms (see Fig.~\ref{fig5}).   

%
On the other hand, when \emph{ intrinsic} parameters are varied, the complex and real overlaps largely agree, even though the complex overlap has more
information available; see  Fig.~\ref{fig.realversuscomplex2}.


%
For the figures and discussion in this paper, we compare the real and complex \emph{normalized} overlaps, dividing each
network response by the network signal to noise.      
For the two identical detectors used in the complex overlap, the network SNR is just a quadrature sum of two detecters; see Eq.~(\ref{eq.realoverlap}), 
\be
\rho^2=\rho^2_+ + \rho^2_-.
\ee
By contrast, for the real overlap, the relevant SNR is just $\rho_+^2$.    For an identical source, a two-detector
network has higher overall SNR.

One point of this paper is to compare the real and complex overlaps from a parameter estimation point of view.
Starting from  Eq.~(\ref{eq.preal}),  the  real overlap enters directly into the expression for the posterior.   As a
result, contours of the  real ambiguity function $(h_0|h)$  should closely correspond to contours of
the posterior parameter distribution, for  measurements limited to a single polarization and known sky location.
By contrast,   for a source with known sky location seen by  a network with comparable sensitivity to both
polarizations, the \emph{complex} overlap enters directly into the expression for the posterior.  
A real network will have unequal sensitivity to two polarizations.  We therefore expect the real posterior will
resemble some average between the posteriors estimated using the single-detector and network overlaps.  
In a subsequent publication we will compare our results with posteriors computed by MCMC, to quantify how well our
simple estimates do at characterizing measurement accuracy.

\section{Reference  frequency and fine-scale structure in the ambiguity function and systematic errors}
\label{ap:ReferenceFrequency}
\begin{figure*}[!]
\centerline{
\includegraphics[width=6cm]{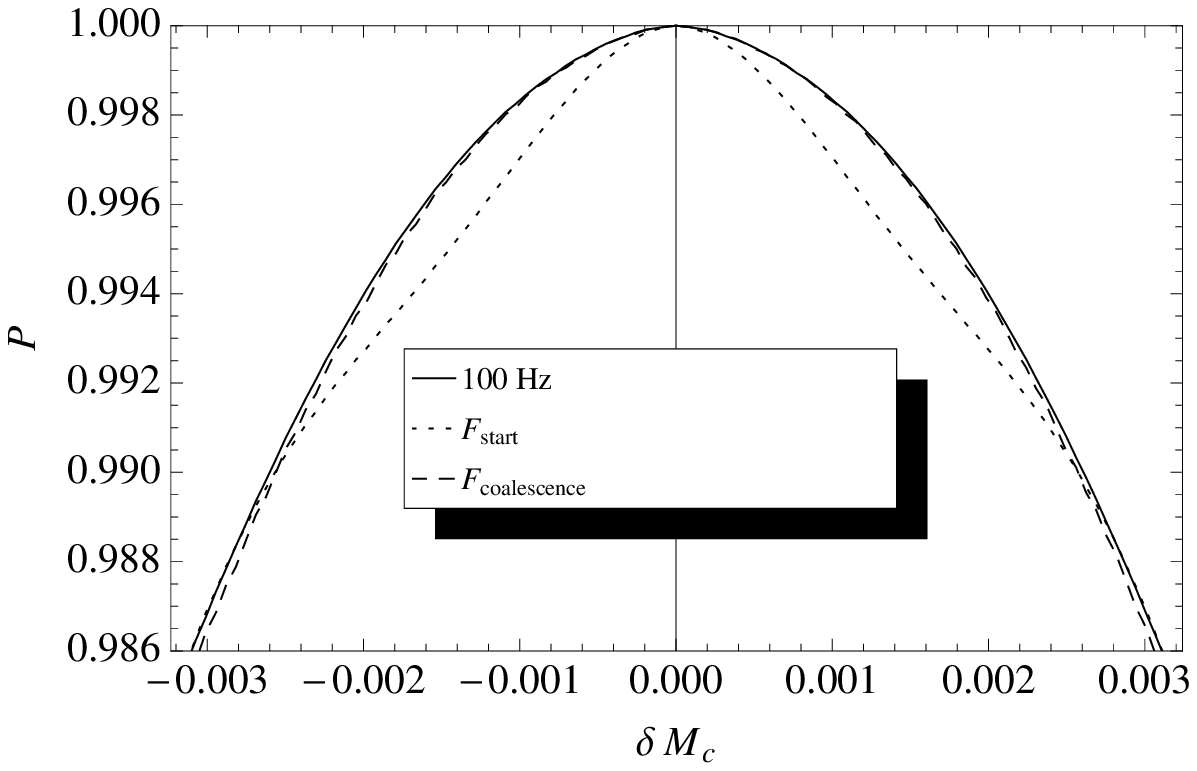}
\includegraphics[width=6cm]{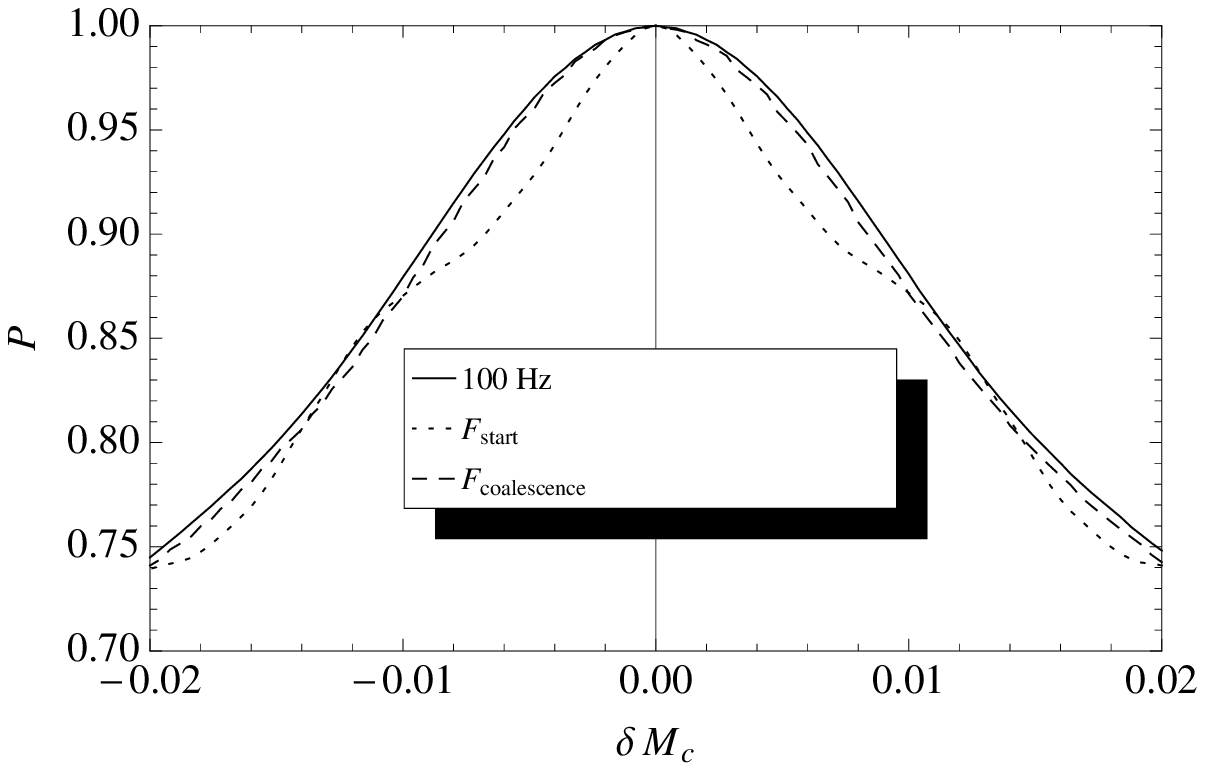}
\includegraphics[width=6cm]{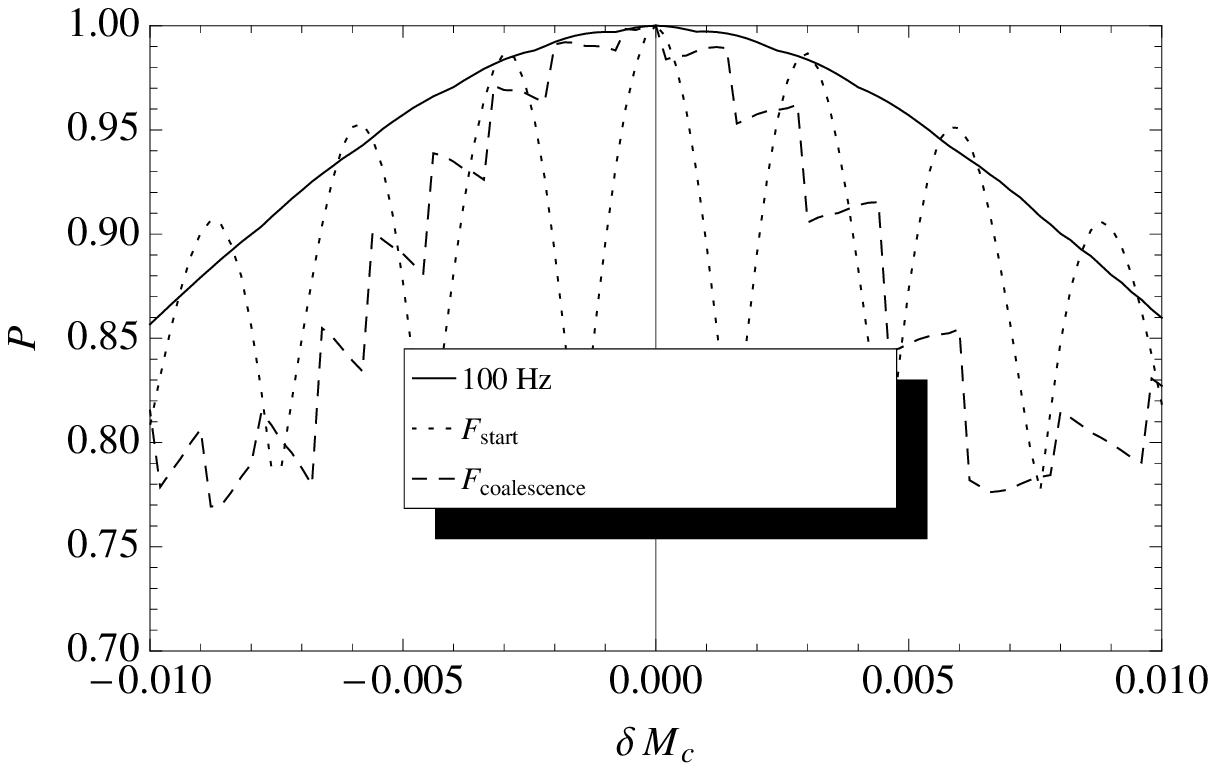}
 }
\caption{\label{fig:AmbiguityAndReference:1d:Angle}{\bf Examples for a dependence of the ambiguity function on the reference frequency.} Left: leading-order waveforms, $\iota=\pi/4$, Middle: higher-order waveforms, $\iota=\pi/4$, Right: leading-order waveforms, $\iota=\pi/2$.} 
\end{figure*}

\begin{figure*}[!]
 \centerline{
\includegraphics[width=6cm]{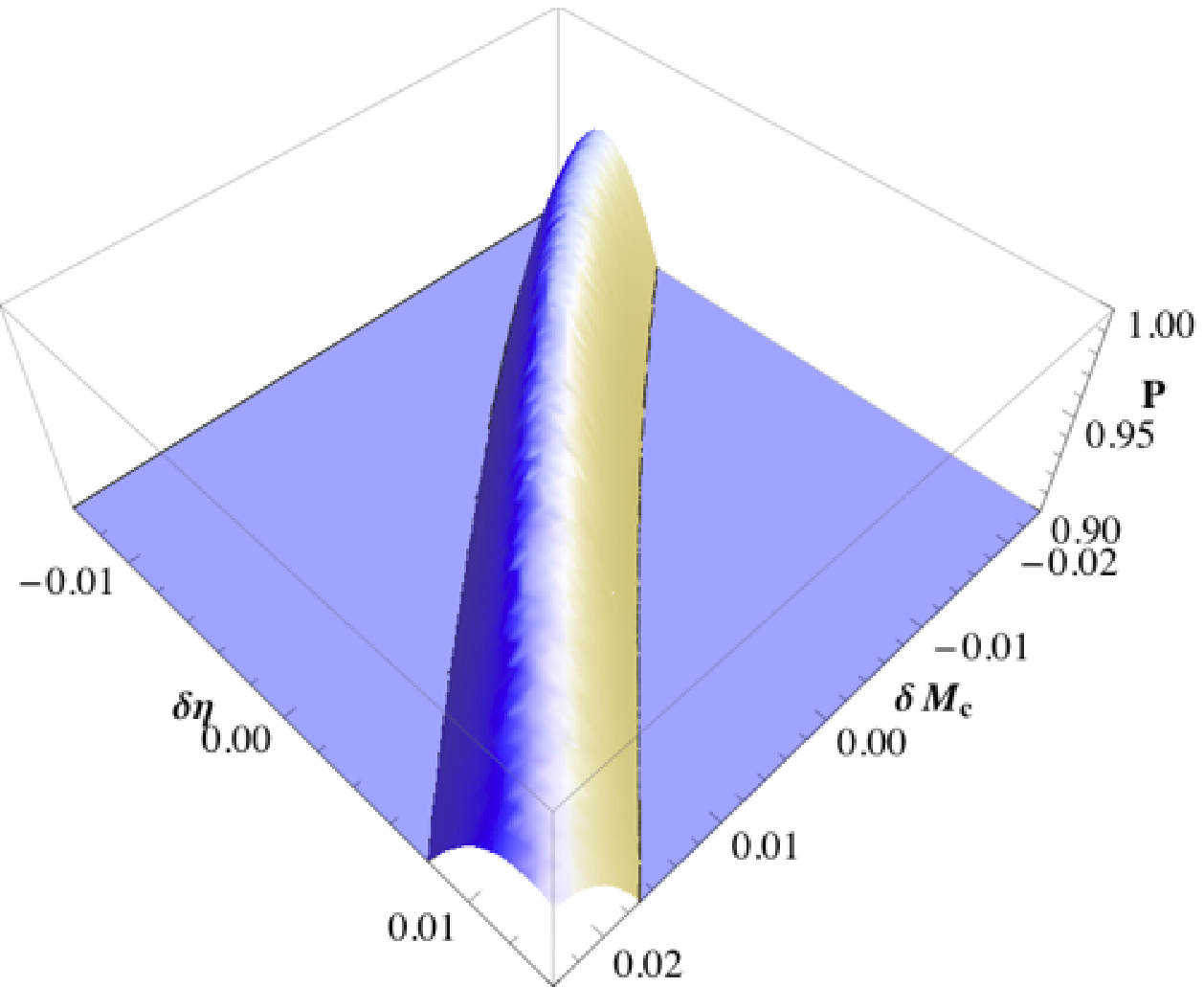}
\includegraphics[width=6cm]{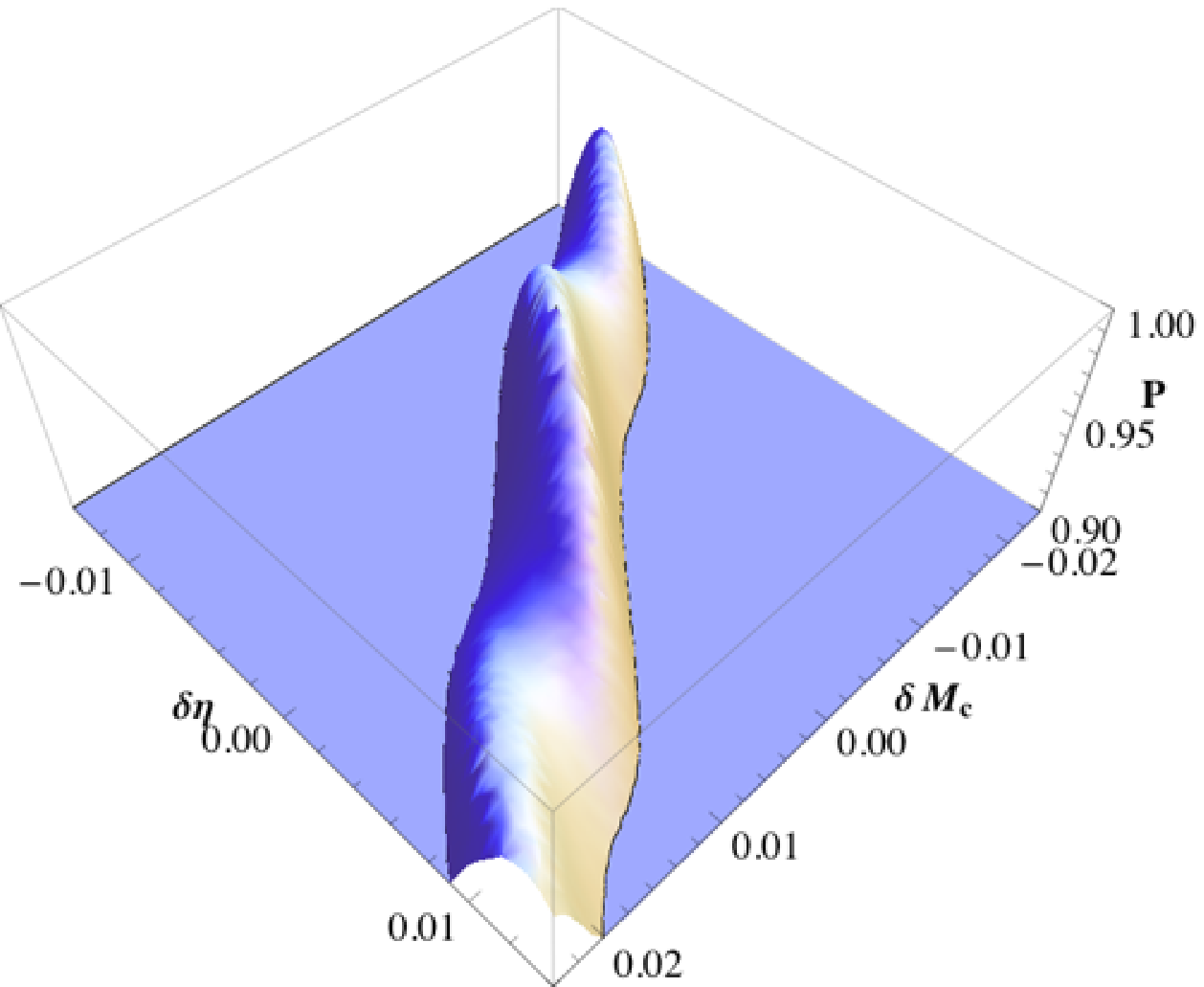}
\includegraphics[width=6cm]{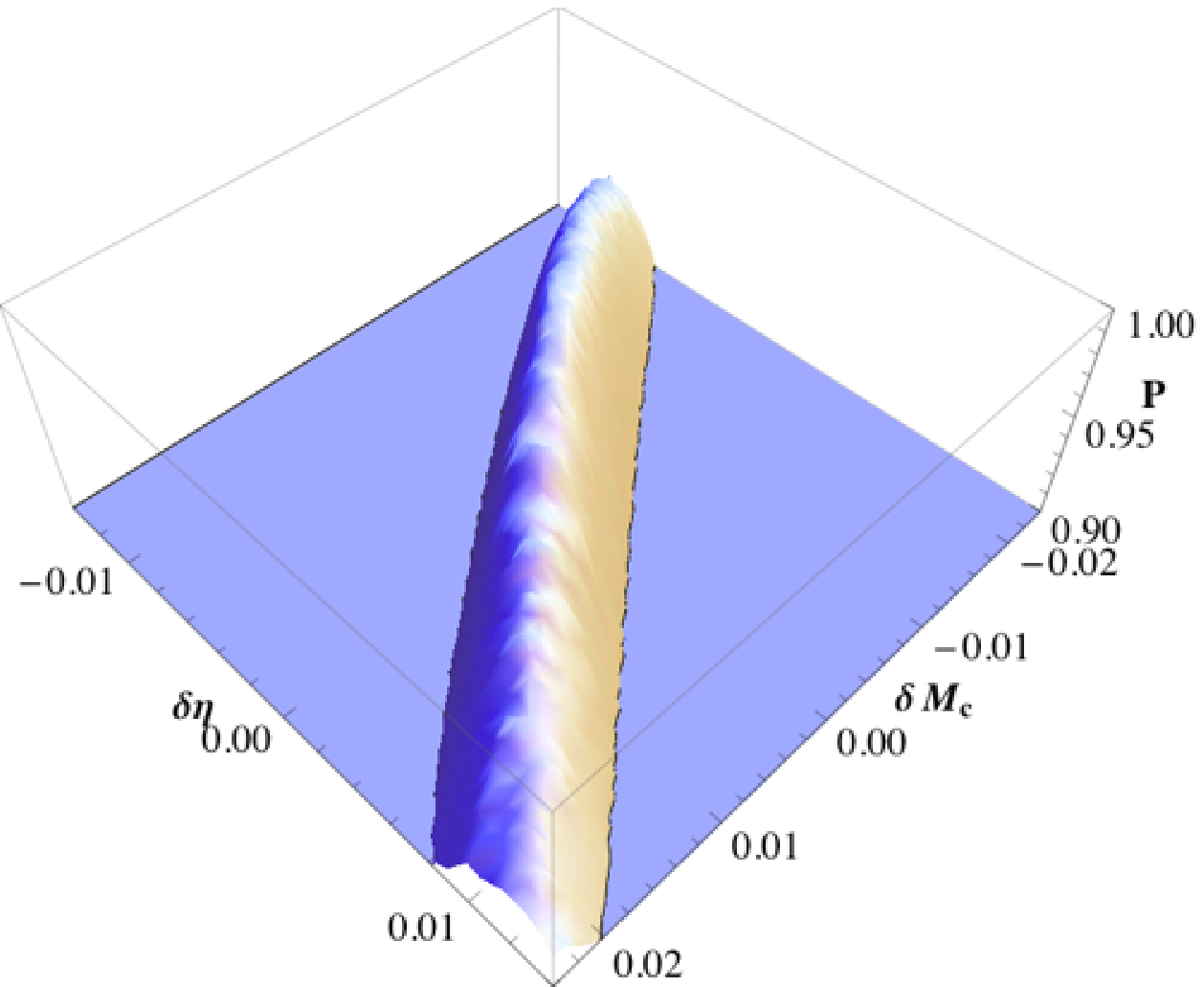}
 }
\caption{\label{fig:AmbiguityAndReference:2d:McEta}{\bf  Examples for a dependence of the ambiguity function on the reference frequency for the higher-order waveforms.} Left: $f_{\rm ref} = 100$ Hz, Middle: $f_{\rm ref} = 40$ Hz (start frequency), Right: $f_{\rm ref} \sim 520$ Hz (coalescence frequency).}  
\end{figure*}

Even for a zero-spin binary with only leading-order harmonics, the choice of reference frequency at which the parameters are defined significantly influences
the structure of the ambiguity function  $P(\lambda_0,\lambda)$.   As concrete examples,   Figs. \ref{eq.ambiguity},
and \ref{fig:AmbiguityAndReference:2d:McEta} show that specifying the orbital phase at the start or end of the waveform
introduces additional structure on physical scales into the ambiguity function for the most well-
determined and physical parameters, the chirp mass $M_{\rm c}$ and mass ratio $\eta$.  
Even more troubling, Fig.~\ref{fig:AmbiguityAndReference:1d:Angle} shows that an ill-chosen reference frequency can
introduce \emph{extremely} fine-scale structure into the ambiguity function, for orientations away from $\hat{z}$.
By contrast, a reference frequency  $f_{\rm ref}$ close to the half-power point of the detector reduces these effects, where
$f_{\rm ref}$ is estimated by 
\begin{eqnarray}
\frac{d\rho^2}{df} &\equiv& 4 \frac{|h(f)|^2}{S_h} \\
\int_0^{f_{\rm ref}} \frac{d\rho^2}{df} &=& \int_{f_{\rm ref}}^{\infty} \frac{d\rho^2}{df}
\end{eqnarray}
for $|h(f)|\propto f^{-7/6}$ the standard restricted amplitude.
What introduces this severe dependence for such a vanilla waveform?  The significant accumulation of orbital and hence
waveform phase between the coordinates' base point and the detector's sensitive band.  Roughly speaking, the detector is
 sensitive to the configuration of the binary as it crosses through its sensitive band.   This trajectory can be
 characterized by some instantaneous parameters $\bar{\lambda}$.  By contrast, the waveform at a significantly earlier
 or later time has rotated roughly $\Delta \Phi/\pi$ times between that frequency and the observed one.  As a result,
 derivatives of the waveform relative to $\lambda$ differ from derivatives relative to $\bar{\lambda}$ by a term of
 order $h \partial \delta \Phi/\partial \bar{\lambda}^a$.   
As a result, the more sensitive the waveform is to a parameter, the more severe the \emph{absolute} impact of adopting a poor reference
frequency.    
%
%

While we have chosen a single reference frequency, we have not optimized it, as we anticipate no single choice will work
well for all parameters.  
The early part of the inspiral carries far more information about the chirp mass and mass ratio, as these both impact
the total number of cycles.   For example, to leading-order $\Gamma_{M_{\rm c} M_{\rm c}} \simeq \left<(\partial_{M_{\rm c}} \Psi)^2 \right>$
where the average $\left< \right>$ corresponds to averaging in signal power (i.e., averaging over a distribution
$\propto d \rho^2/df$).  Evaluating the derivative and writing as an integral over frequency, we see the corresponding
integrand to $d\rho^2/df$ in the definition of  $\Gamma_{M_{\rm c}M_{\rm c}}$ is weighted by a polynomial in $f$ that is
\emph{heavily} biased towards low frequencies:
\begin{eqnarray}
\Gamma_{M_{\rm c} M_{\rm c}} \propto \int [f^{-10/3}+ \text{lower powers}] d\rho^2/df
\end{eqnarray}
By contrast, spin effects enter at $v^3 \propto f$ past leading-order in phase, suggesting the spin-spin Fisher matrix
components are less severely biased towards low frequency;
\begin{eqnarray}
\Gamma_{SS} \propto \int f^{-10/3 + 2} d\rho^2/df
\end{eqnarray} 
Other components of the Fisher matrix arise from other averages over frequency and have other break-even points.  In
other words, for each measurable quantity, a generally distinct  epoch of the signal provides the most information.  No
one breakeven point works for all.


\end{document}